\documentclass{aa}
\usepackage{natbib}
\bibpunct{(}{)}{;}{a}{}{,} 
\usepackage{graphicx}
\graphicspath{{Images/}}
\usepackage{txfonts}
\usepackage{subfig}
\usepackage{array}
\usepackage{orcidlink}
\usepackage{hyperref}
\hypersetup{
    colorlinks=true,
    linkcolor=blue,     
    urlcolor=blue,
    citecolor=blue
    }

\def\NH{$N$\textsubscript{H}}
\def\NHeq{$N$\textsubscript{H,eq}}

\newcommand{\customcite}[2]{%
  \citeauthor{#1}\ (\citeyear{#1}; #2)%
}

\begin{document} 

\title{Self-consistent population synthesis of AGN from observational constraints in the X-rays}


\author{Dimitra Gerolymatou\inst{1}\orcidlink{0009-0004-8853-1718},
          Stéphane Paltani\inst{1}\orcidlink{0000-0002-8108-9179},
          Claudio Ricci\inst{2,3}\orcidlink{0000-0001-5231-2645},
          Manon Regamey\inst{1}\orcidlink{0009-0008-2709-121X}
          }

\institute{Department of Astronomy, University of Geneva, 1290 Versoix, Switzerland \and Instituto de Estudios Astrofísicos, Universidad Diego Portales, Chile \and Kavli Institute for Astronomy and Astrophysics, Peking University, Beijing 100871, China}


 
\abstract
   {The cosmic X-ray background (CXB) is produced by the emission of unresolved active galactic nuclei (AGN), thus providing key information about the properties of the primary and reprocessed X-ray emission components of the AGN population. Equally important, studies of individual sources provide additional constraints on the properties of AGN, such as their luminosity and obscuration.
   Until now, these constraints have not been self-consistently addressed by intrinsically linking emission, absorption, and reflection. Here we perform numerical simulations with the ray-tracing code, \textsc{RefleX}, which allows us to self-consistently model the X-ray emission of AGN with flexible geometries for the circumnuclear medium.
   Using the \textsc{RefleX}-simulated emission of an AGN population, we attempt to simultaneously reproduce the CXB and absorption properties measured in the X-rays, namely the observed fraction of \NH{} in bins of log(\NH{}) and the fraction of absorbed AGN, including their redshift and luminosity evolution. We sample an intrinsic X-ray luminosity function and construct gradually more complex physically motivated geometrical models.
   We examine how well each model can match all observational constraints using a simulation-based inference (SBI) approach.
   We find that, while the simple unification model can reproduce the CXB, a luminosity dependent dusty torus is needed to reproduce the absorption properties. When adding an accretion disc, the model best matches all constraints simultaneously. Our synthetic population is able to reproduce the dependence of the covering factor on luminosity, the AGN number counts from several surveys, and the observed correlation between reflection and obscuration. Finally, we derive an intrinsic Compton-thick fraction of 21$\pm$7\%, consistent with local observations.}


\titlerunning{Self-consistent AGN population synthesis in the X-rays}
\authorrunning{D. Gerolymatou et al}
\maketitle

\section{Introduction} \label{sec:intro}
Supermassive black holes (SMBHs) at the cores of galaxies are the engines of active galactic nuclei (AGN) during an accreting phase. The SMBHs are closely linked to essential properties of their host galaxies, including mass, velocity dispersion, and luminosity \citep[e.g.,][]{1998AJ....115.2285M, 2000ApJ...539L..13G, 2013ARA&A..51..511K}. Such correlations hint at the potential for SMBHs to regulate star formation rates and play a significant role in galaxy evolution. To comprehensively probe the evolution of SMBHs, AGN demographics have been examined through population studies, allowing to build their luminosity function \citep[e.g.,][]{2014ApJ...786..104U, 2015MNRAS.451.1892A, 2019ApJ...871..240A} and other properties such as the black hole mass function and Eddington ratio distribution function. 

X-ray radiation is a universal feature of AGN, which typically outshines the emission of the host galaxy \citep[e.g.,][]{2015A&ARv..23....1B}. Furthermore, X-ray observations are a very efficient method for detecting obscured AGN as X-ray photons can penetrate the high column density of the surrounding dusty material, thus allowing us to study the innermost regions of AGN. The primary X-ray emission is attributed to the inverse Comptonisation of accretion disc photons in a hot corona close to the SMBH \citep[e.g.,][]{1991ApJ...380L..51H}, while reprocessed components come from reflection on the circumnuclear material \citep[e.g.,][]{1990Natur.344..132P, 1994MNRAS.267..743G}. The circumnuclear material is thought to be in the form of a dusty torus (see \citealp{2018ARA&A..56..625H} for an overview) and acts as an absorber and reflector at the same time, while other dense structures such as the accretion disc can also act as reflector. The integrated emission from the AGN population is imprinted on the cosmic X-ray background (CXB; e.g., \citealp{2022hxga.book...78B}). A clear peak in the CXB spectrum is observed between $20-30$\,keV, which is attributed to the X-ray radiation reflected on the circumnuclear material. The CXB is often used as one of the constraints for AGN population synthesis models \citep[e.g.,][]{2007A&A...463...79G, 2012A&A...546A..98A, 2014ApJ...786..104U, 2019ApJ...871..240A}; however, it can be reproduced by different AGN populations, for instance by assuming different AGN spectral features, shape of the X-ray luminosity function (XLF), intrinsic neutral hydrogen column density (\NH{}) distribution, and fraction of Compton-thick (CT) AGN (which we define as the fraction of AGN with \NH{} $\geq 10^{24}$ cm$^{-2}$), as demonstrated in \cite{2009ApJ...696..110T}, \cite{2012A&A...546A..98A} and \cite{2019ApJ...871..240A}. Some degeneracy between these parameters could be lifted by using AGN number counts obtained in X-ray surveys \citep[e.g.,][]{2014ApJ...786..104U, 2015MNRAS.451.1892A, 2019ApJ...871..240A}.

X-ray surveys of AGN also offer insights into their absorption properties, which strongly affect the emission from AGN and thus the modelling of the CXB. The fraction of absorbed AGN (usually defined as the fraction of AGN with \NH{} $\geq 10^{22}$ cm$^{-2}$) has been consistently found to be anti-correlated with luminosity \citep[e.g.,][]{1991MNRAS.252..586L, 2003ApJ...598..886U, 2008A&A...490..905H, 2014ApJ...786..104U, 2015MNRAS.451.1892A, 2015ApJ...802...89B}, suggesting that there are intrinsic physical differences between the populations of obscured and unobscured AGN. The intrinsic fraction of CT AGN is still debated, with previous studies finding values from 10 to 50$\%$ \citep[e.g.,][]{2011MNRAS.413.1206B, 2012A&A...546A..98A, 2014ApJ...786..104U, 2015ApJ...815L..13R, 2019ApJ...871..240A}. As the CXB peak at $20-30$\,keV can be reproduced by either a high fraction of CT AGN or stronger reflected X-ray emission from all AGN, the amount of reflection must be anti-correlated with the fraction of CT AGN. Studies with stronger reflection components find a CT fraction as low as 10$\%$ \citep[e.g.,][]{2009ApJ...696..110T, 2011A&A...532A.102R}, while the most recent population synthesis of AGN in \cite{2019ApJ...871..240A}, which is able to reproduce simultaneously the CXB and AGN number counts from multiple X-ray surveys, predicts an intrinsic CT fraction of 56$\pm$7\% for $z\lesssim1$.

An important caveat of all previous studies is that the absorption of AGN is assumed and is not physically linked with any other AGN properties. Until now, X-ray observables concerning the absorption properties of AGN, such as the observed column density distribution and the dependence of the absorbed fraction on X-ray luminosity and redshift, have not been self-consistently addressed by physically linking obscuration and reflection. In this study, we create a fully self-consistent population synthesis model of AGN, where the X-ray obscuration and reflection are linked through the distribution of matter. In order to do that, we carry out numerical simulations using the ray-tracing code \textsc{RefleX} \citep{2017A&A...607A..31P}, which allows us to simulate the X-ray emission from AGN with different geometries. We sample an intrinsic X-ray luminosity function to create a synthetic AGN population and use gradually more complex physically motivated geometrical models. With each model, we attempt to reproduce the CXB and the observed absorption properties. We use a simulation-based inference approach \citep{tejero-cantero2020sbi, 2019arXiv190507488G} to compare the models to the observations and infer the model parameters that best match all constraints simultaneously. Finally, we examine the luminosity dependence of absorption, the AGN number counts at different energy bands, the reflection-obscuration correlation, the impact of galactic obscuration on our population, and derive the intrinsic CT fraction of the AGN population. Throughout this paper, all logarithmic values (log) refer to logarithms with base 10, and we adopt the following cosmological parameters: $H$\textsubscript{0} = 70\,km\,s$^{-1}$\,Mpc$^{-1}$, $\Omega$\textsubscript{m} = 0.3, and $\Omega$\textsubscript{$\Lambda$} = 0.7.

\section{X-ray observational constraints} \label{sec:observations}
\begin{figure*}[t]
\centering
\subfloat{\includegraphics[width=0.5\hsize]{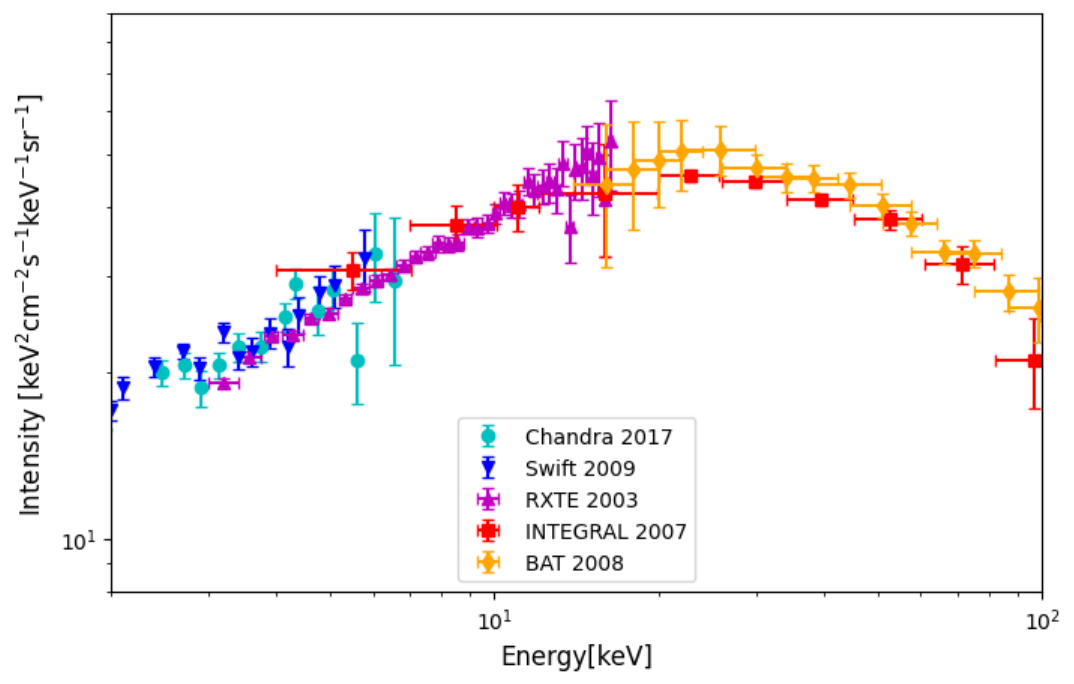}} \\\subfloat{\includegraphics[width = 0.314\hsize]{Images/Nhdist.pdf}}
\subfloat{\includegraphics[width = 0.33\hsize]{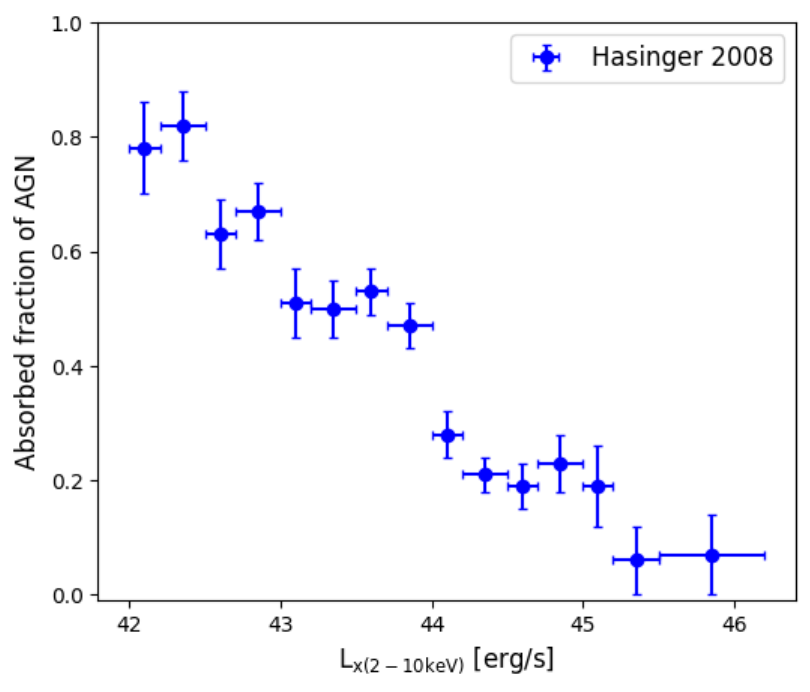}}
\subfloat{\includegraphics[width = 0.332\hsize]{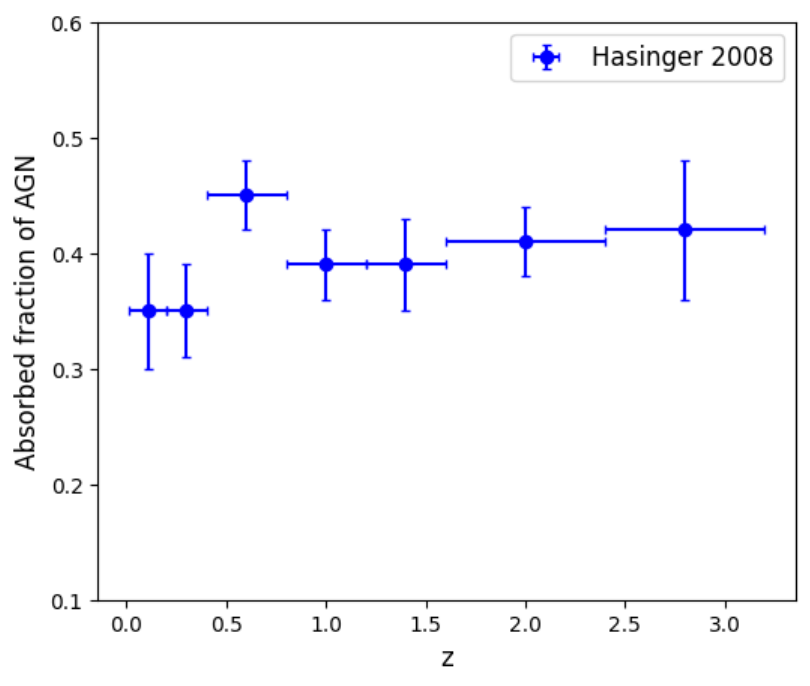}} 
\caption{X-ray observables used as constraints in this work: \textit{Top:} CXB measured by various instruments: Chandra \citep[cyan;][]{2017ApJ...837...19C}, INTEGRAL \citep[red;][]{2007A&A...467..529C}, RXTE \citep[magenta;][]{2003A&A...411..329R}, \textit{Swift}/BAT \citep[orange;][]{2008ApJ...689..666A} and \textit{Swift}/XRT \citep[blue;][]{2009A&A...493..501M}. \textit{Bottom left:} observed \NH{} distribution of 731 AGN in the 70-month \textit{Swift}/BAT catalog, detected in the 14--195\,keV range, from \cite{2017ApJS..233...17R}. The point with the dashed error bars represents the sum of the first two bins, used in this work (see \ref{subsec:obscurationobs} for more information). \textit{Bottom centre:} absorbed fraction of 1290 AGN from multiple surveys in the 2--10\,keV range as a function of observed luminosity from \cite{2008A&A...490..905H}. \textit{Bottom right:} same absorbed fraction as a function of redshift. \label{fig:observations}}
\end{figure*}

To constrain the geometrical and physical properties of the AGN population, we need to gather several observational AGN properties. In Fig. \ref{fig:observations}, we show the X-ray observations used as constraints in this work, and we describe each of them in this section. The constraints are the CXB and several absorption properties from studies of X-ray surveys, namely the observed \NH{} distribution and the dependence of the observed fraction of absorbed AGN on observed X-ray luminosity and on redshift. We choose observed properties because the intrinsic ones depend on the assumptions made in other papers. Instead, we apply the selection function of each survey on our synthetic population.

\subsection{Cosmic X-ray background} \label{subsec:CXB}
The CXB is formed by the integrated emission of all AGN across cosmic time (see \citealp{2022hxga.book...78B} for an overview). As our first constraint, we use some of the most recent measurements of the CXB, comprehensively covering our target energy range, obtained from five instruments: Chandra, INTEGRAL (JEM-X, IBIS/ISGRI, SPI), RXTE/PCA, \textit{Swift}/BAT and \textit{Swift}/XRT (\citealp{2017ApJ...837...19C, 2007A&A...467..529C, 2003A&A...411..329R, 2008ApJ...689..666A,2009A&A...493..501M}; Fig. \ref{fig:observations} top). All measurements find the CXB spectral index to be $\sim1.4$ below 10\,keV. Any normalisation discrepancies between data sets may be mostly the result of inaccurate cross-calibration of spectra, unclear instrumental backgrounds, and cosmic variance \citep[e.g.,][]{2009A&A...493..501M}. We note the pioneering role of the HEAO-1 mission in the discovery and early characterisation of the CXB shape \citep{1999ApJ...520..124G}. The statistical disagreement in normalisation between HEAO-1 and our chosen measurements is well documented and discussed in the literature \citep[see for example][and references therein]{2007A&A...467..529C, 2008ApJ...689..666A, 2009A&A...493..501M}.

We consider the energy range 2--100\,keV, as at $E$ > 100\,keV there is a significant contribution from blazars \citep[e.g.,][]{2009ApJ...707..778D}, and at $E <$ 2\,keV the emission of AGN becomes quite complex because of absorption, soft-excess emission, and non-negligible contribution from galaxies \citep[e.g.,][]{2012MNRAS.427..651C}.

\subsection{Absorption properties} \label{subsec:obscurationobs}

\subsubsection{Column density distribution}
The second constraint is the observed line-of-sight \NH{} distribution of local (median redshift $z$ = 0.0367) AGN from \customcite{2017ApJS..233...17R}{hereafter R17}. Their sample includes 731 non-blazar AGN detected in the 14--195\,keV energy band in the \textit{Swift}/BAT 70-month all-sky survey. Being less biased against high \NH{} AGN, it consists of $\sim 50 \%$ obscured AGN. The hard X-ray sensitivity of BAT allows us to investigate the distribution of absorption up to the CT regime, albeit with significant selection bias against CT AGN.

The \NH{} distribution presented in R17 shows some evidence of bimodality with the highest peak for log(\NH{}/\,cm$^{-2})\leq$ 21 and the secondary in the log(\NH{}/\,cm$^{-2}$) = 23--24 bin (see Fig. \ref{fig:observations} bottom left). The R17 distribution is largely consistent with NuSTAR observations in the 8--24\,keV band, which also show two peaks at the same \NH{} \citep[e.g.,][]{2018ApJ...854...33Z}. This bimodality suggests that the obscuration may be caused by different structures, possibly due to the plane of the galaxy hosting the AGN at lower \NH{} and due to the denser circumnuclear material at higher \NH{} \citep[e.g.,][]{2008A&A...485..707P}. Considering the purpose of this study, we do not model the host galaxy, resulting in a lack of mildly obscured (\NH{} $< 10^{22}$ cm$^{-2}$) AGN. Hence, we compare the number of simulated unobscured objects to all AGN with \NH{} $< 10^{22}$ cm$^{-2}$, assuming galactic column densities log($N$\textsubscript{H}/\,cm$^{-2}$) $< 22$ \citep[e.g.,][]{2009MNRAS.400.2050G}.

\subsubsection{Fraction of absorbed AGN}
The last two constraints for this work are the observed fraction of absorbed AGN as a function of the observed X-ray luminosity and as a function of redshift from \customcite{2008A&A...490..905H}{hereafter H08}. In H08, 1290 AGN detected in the 2--10\,keV range from ten independent samples (summarised in table 1 in their paper) are classified into type-I (unabsorbed) or type-II (absorbed) by combining optical spectroscopy and X-ray hardness ratios. Optical/UV broad lines are used to classify sources into broad-line AGN and narrow-line AGN. The distinction found between the two classes corresponds to \NH{} $\approx 3 \times 10^{21}$ cm$^{-2}$, above which only narrow lines are observable. Thus for the calculation of the absorbed fraction, we consider as absorbed the AGN with log(\NH{}/\,cm$^{-2}$) $ >$ 21.5 [instead of the usual log(\NH{}/\,cm$^{-2}$)$ > 22$] for a meaningful comparison with the H08 results.

In H08, the fraction of absorbed AGN is found to decrease with increasing logarithmic luminosity with a slope of $-0.226 \pm 0.014$ (fitted with a linear function) for log($L$\textsubscript{X}/ erg s$^{-1}$) $\geq$ 42. The presence of the anti-correlation has been confirmed by multiple studies carried out in the last three decades \citep[e.g.,][]{1991MNRAS.252..586L, 2003ApJ...598..886U, 2014ApJ...786..104U, 2015MNRAS.451.1892A, 2015ApJ...802...89B}. At the same time in H08, the fraction of absorbed AGN is found to increase with redshift as $(1+z)^{0.48}$ over the redshift range 0--3.2 at fixed luminosity. Such evolution has been confirmed in later studies \citep{2014ApJ...786..104U,2015MNRAS.451.1892A,2017ApJS..232....8L,2020A&A...639A..51I}. Here, however, we consider the fraction integrated over the entire range of luminosity to avoid larger uncertainties caused by the low number of sources when the samples are divided into luminosity bins. In this case, the dependence of the fraction of absorbed AGN on redshift is flat (Fig. \ref{fig:observations} bottom right).

We opt for H08 instead of the more recent works cited above because it provides the absorbed fraction across broad luminosity and redshift ranges, in addition to using observed X-ray luminosities. This reduces the uncertainties that can arise from dividing in narrow bins with limited detected sources and avoids the additional assumptions made in other studies to infer intrinsic luminosities. Moreover, this is a consistent set of constraints (as some studies do not provide the absorbed fraction as functions of both luminosity and redshift) and sufficient (as we do not expect that our simplistic models can explain all detailed features). Finally, in contrast to R17, the H08 sample contains both local and high redshift AGN, meaning that our synthetic population has to be able to simultaneously match the absorption properties of AGN detected at different redshifts as well as in different energy ranges (2--10\,keV and 14--195\,keV).


\section{Modelling of AGN population} \label{sec:modelling}
We seek to create a self-consistent synthetic population of AGN that can simultaneously reproduce the CXB and the observed absorption properties of AGN. With our work we look for the geometrical model that can best match the observables. In this section, we present the tools, models, and assumptions that we use to create the synthetic AGN population and simulate their emission.

\subsection{RefleX} \label{subsec:reflex}

We use the ray-tracing code \textsc{RefleX}\footnote{\url{https://www.astro.unige.ch/reflex/}} \citep{2017A&A...607A..31P} to self-consistently simulate the X-ray emission of AGN with different geometries. \textsc{RefleX} performs realistic propagation of X-ray (0.1-1 MeV) photons through the circumnuclear material of AGN, using Monte Carlo methods. Arbitrary geometries with spatially varying composition and density can be selected by the user. Moreover, the geometry and emission spectrum of the X-ray source can also be selected. The most important physical processes in the X-rays are implemented, namely Compton and Rayleigh scattering, photoelectric absorption, and fluorescence. In \textsc{RefleX} 3.0 \citep{2023ApJ...945...55R} dust absorption and scattering is also included following the Milky Way model from \cite{2003ApJ...598.1026D}. We use the latest version of \textsc{RefleX} to generate spectra at specific inclinations and calculate fluxes in selected energy bands.

\subsection{X-ray luminosity function} \label{subsec:luminosityfunction}
To create a population of AGN for our simulations, we need an X-ray luminosity function, which provides the AGN number density in Mpc$^{-3}$ dex$^{-1}$ as a function of redshift and logarithmic X-ray luminosity. We choose the intrinsic (i.e. absorption corrected) rest-frame 2--10\,keV luminosity function derived in \customcite{2003ApJ...598..886U}{hereafter U03}, frequently adopted in previous population syntheses models of the CXB \citep[e.g.,][]{2007A&A...463...79G, 2012A&A...546A..98A, 2016A&A...590A..49E}. The U03 XLF is built from a sample containing 247 HEAO-1, ASCA and Chandra detected sources over the 2--10\,keV range, thus they assume that there is a negligible percentage of CT AGN in their sample.

We opt for U03 instead of more recent studies where the XLFs are either non-parametric \citep[e.g.,][]{2015ApJ...802...89B, 2019ApJ...871..240A}, derived separately for each AGN type \citep[e.g.,][]{2015MNRAS.451.1892A}, or their sample contains CT AGN \citep[e.g.,][]{2014ApJ...786..104U}. As shown in \cite{2014ApJ...786..104U} and \cite{2019ApJ...871..240A}, the log(\NH{}/\,cm$^{-2}) < 24$ XLFs mostly agree with U03, even though they assume different \NH{} distribution functions.

Our synthetic AGN population automatically creates a number of CT AGN. Therefore, our LF is normalised so that the number density of AGN with \NH{} $< 10^{24}$ cm$^{-2}$ matches the U03 XLF. The fraction of CT AGN is then determined self-consistently from the model parameters obtained from the posterior distributions. It is therefore an advantage to use an XLF derived from a sample containing a negligible amount of CT AGN, especially considering that X-ray surveys are biased against detecting heavily CT AGN even at >10\,keV.

The assumptions made in U03 (or later XLFs) will affect our parent distribution and thus our results. Although it would be possible to include the fitting of the parameters of the XLF, for the purposes of this work we use the best-fit values derived in Table 3 of U03. In their work, the XLF, $\phi(L\mathrm{_X^{int}}, z)$, is described by a luminosity-dependent density evolution (LDDE), starting with a smoothly connected double power-law at z = 0:
\begin{equation}
\phi(L\mathrm{_X^{int}}, z=0) = \frac{\mathrm{d}N(L\mathrm{_X^{int}}, z=0)}{\mathrm{d}\mathrm{log}L\mathrm{_X^{int}}} = K \left[\left(\frac{L\mathrm{_X^{int}}}{L_*}\right)^{\gamma_1} + \left(\frac{L\mathrm{_X^{int}}}{L_*}\right)^{\gamma_2}\right]^{-1},
\end{equation}
where $K$ is the normalisation in units of $10^{-6}$\,Mpc$^{-3}$, $L_*$ the characteristic break luminosity in erg\,s$^{-1}$, $\gamma_1$ and $\gamma_2 $ the faint-end and bright-end slopes respectively. The XLF evolves as a function of redshift and luminosity:
\begin{equation}
\phi(L\mathrm{_X^{int}}, z) = \phi(L\mathrm{_X^{int}}, z=0)~e(z,L\mathrm{_X^{int}}).
\end{equation}
The evolution term, $e(z,L\mathrm{_X^{int}})$, is described by a power-law function of (1+$z$) with a cut-off redshift which is a function of luminosity, as defined in Eq. 16 and 17 in U03.

\subsection{Models} \label{subsec:models}

\begin{table*}[t]
	\begin{center}
    	\caption{Overview of the models explored, their free parameters, and priors as discussed in Sect. \ref{subsec:models}.} \label{tab:models}
		\begin{tabular}{l c c c c c}
        \hline\hline
			\rule{0pt}{1em}Model\tablefootmark{1} & $\mu$\,(log/\,cm$^{-2}$)\,\tablefootmark{2} & $\sigma$\,(log/\,cm$^{-2}$)\,\tablefootmark{2} & $R_0$\tablefootmark{3,a} & $\alpha$\,\tablefootmark{3} & $R_1$\tablefootmark{4,a}\\
			\hline
			\rule{0pt}{1em}Simple torus & 21.5--25.5 & 0.01--2.5 & - & - & 0.01--1.5 \\
			LD torus & 21.5--25.5 & 0.01--2.5 & 0.01--1.5 & 0.01--1.0 & - \\
			LD+AD & 21.5--25.5 & 0.01--2.5 & 0.01--1.5 & 0.01--1.0 & - \\
        \hline 
		\end{tabular}
        \tablefoot{\tablefoottext{1}{Simple torus, luminosity dependent (LD) torus, and LD torus with an accretion disc (AD).}
        \tablefoottext{2}{Mean, $\mu$, and standard deviation, $\sigma$, of the tori \NHeq{} distribution.}
        \tablefoottext{3}{Normalisation, $R_0$, and slope, $\alpha$, in Eq. (\ref{eq:RoutLx}).}
        \tablefoottext{4}{Scaling parameter, $R_1$, of the simple torus.}\\
        \tablefoottext{a}{Arbitrary units for models without AD and measured in pc for the AD model.}}
	\end{center}
\end{table*}

\begin{figure}[t]
\centering
\includegraphics[width=\hsize]{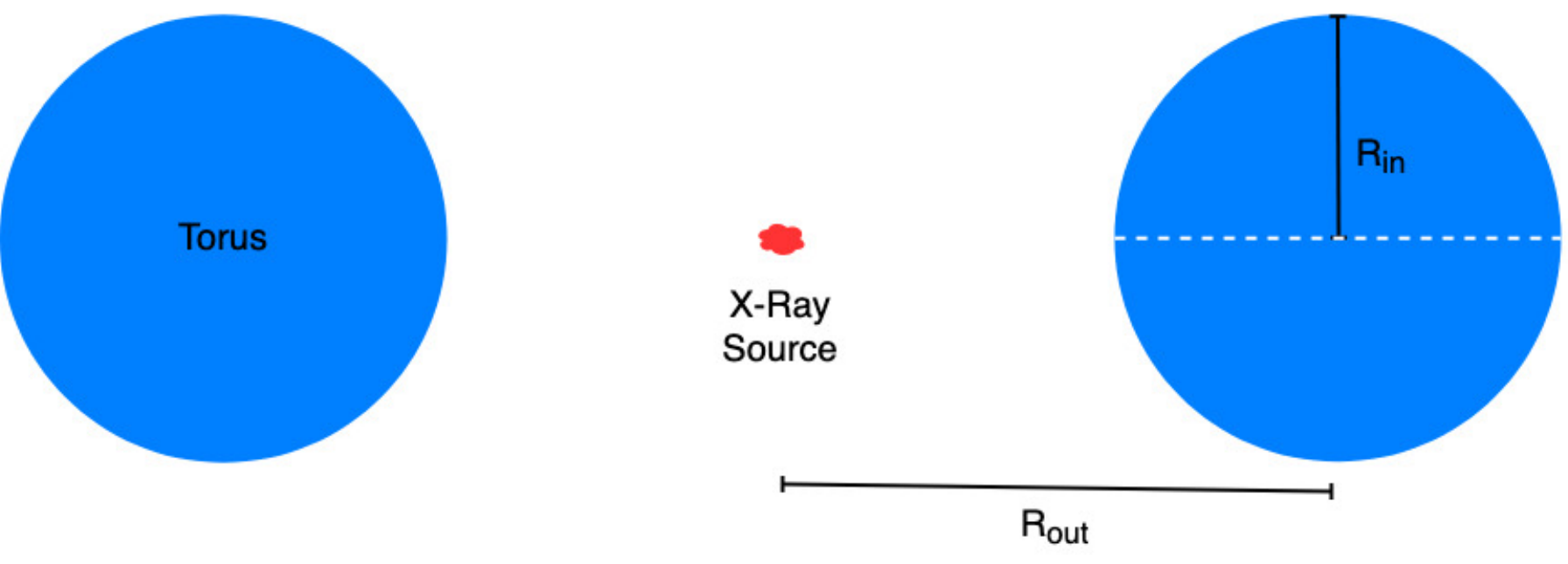}
\caption{Torus geometry in \textsc{RefleX}. The inner radius, $R$\textsubscript{in}, is defined as the radius of the cross-section, while the outer radius, $R$\textsubscript{out}, is measured from the centre of the torus to the centre of the cross-section. The equatorial column density, \NHeq{}, is measured across the diameter of the cross-section (white dashed line). The X-ray point source is placed at the centre of the torus.
\label{fig:torus}}
\end{figure}

The physical models of the AGN population that we consider and their geometries constructed in \textsc{RefleX} are explained below and summarised in Table \ref{tab:models}. All of our models comprise a cold dusty homogeneous torus, the geometry of which is shown in Fig. \ref{fig:torus}. The torus is fully described in \textsc{RefleX} by its inner, $R$\textsubscript{in}, and outer, $R$\textsubscript{out}, radii as well as its equatorial column density, \NHeq{}, and its composition. We set the molecular hydrogen fraction to 0.5 \citep[e.g.,][]{2009ApJ...702...63W}, the dust fraction (defined as the fraction of Fe atoms locked into dust grains) to 1, the matter composition to \cite{2003ApJ...591.1220L}, and assume solar metallicity. As shown in \cite{2022MNRAS.512.2961M}, varying standard abundances do not result in significant differences in the simulated spectra at the level of accuracy we want to achieve here.

For the two models that include only the torus, we place the X-ray point source at the centre of the torus as a good approximation. In all cases, we choose a power-law emission spectrum for the X-ray source with typical values for the photon index, $\Gamma$ = 1.9, and high-energy exponential cutoff, $E$\textsubscript{C} = 200\,keV \citep[e.g.,][]{2017ApJS..233...17R, 2018MNRAS.480.1819R}.

\subsubsection{Simple torus}
We start from the simple unification of AGN, which suggests that the observed differences between type-I and type-II AGN are caused only by orientation effects depending on the observer's line-of-sight \citep{1993ARA&A..31..473A}. Our model is composed of an obscuring torus whose $R$\textsubscript{in}/$R$\textsubscript{out} ratio is a free parameter independent of any other AGN properties and remains the same throughout the population. Hence, in this population the only difference between different AGN is the torus \NHeq{}. Because the geometry of this model is only dependent on the ratio $R$\textsubscript{in}/$R$\textsubscript{out} which sets the covering factor, the simulation can be scale-free. We fix $R$\textsubscript{in} = 0.5 (in arbitrary units) and let $R$\textsubscript{out} vary as $R$\textsubscript{out} = $R$\textsubscript{in} + $R_1$ with $R_1$ a free positive parameter.

\subsubsection{Luminosity dependent torus} \label{subsec:ldtorus}
We choose our next physically motivated model to include a relationship between the location of the inner edge of the torus and the luminosity. Thus, we construct a model based on the "receding torus" \citep{1991MNRAS.252..586L}, which is often invoked to explain the decrease of the fraction of absorbed AGN as a function of luminosity. In fact, it applies to a dusty torus, where the dust is destroyed when it reaches a temperature of $\sim1500$K \citep[see][]{2008ApJ...685..160N} due to irradiation by the central engine. In the receding torus model, the inner edge of the obscuring torus, $r$\textsubscript{dust} = $R$\textsubscript{out} $-$ $R$\textsubscript{in}, is set by the dust sublimation radius, which increases with luminosity as $r$\textsubscript{dust} $\propto L^{0.5}$. Therefore, if dust-free gas behaves like dusty gas, the fraction of absorbed AGN is expected to scale as $\propto$ $L^{-0.5}$ at high luminosities. 

Our luminosity dependent model (hereafter LD) introduces a dependence of the covering factor of the scaleless tori on the intrinsic X-ray luminosity of the AGN, with the assumption that gas behaves similarly to dust. We let the slope of the dependence vary because the 0.5 index of the receding torus model does not agree with observations of the absorbed fraction in the X-rays, which find a much flatter slope \citep[e.g., H08,][]{2011ApJ...728...58B}. We discuss the possible mechanisms for this in Sect. \ref{subsec:absorptiondependence}. For practical reasons, we fix again $R$\textsubscript{in} = 0.5 (in arbitrary units) and let $R$\textsubscript{out} scale with luminosity as:
\begin{equation}\label{eq:RoutLx}
r_\mathrm{dust} = R_\mathrm{out} - R_\mathrm{in} = R_0 \left(\frac{L\mathrm{_X^{int}}}{L_0}\right)^{\alpha},
\end{equation}
where $L$\textsubscript{0} = $10^{44}$\,erg\,s$^{-1}$ is arbitrarily fixed, and $R$\textsubscript{0} and $\alpha$ are free parameters.

\begin{figure}[t]
\centering
\includegraphics[width=\hsize]{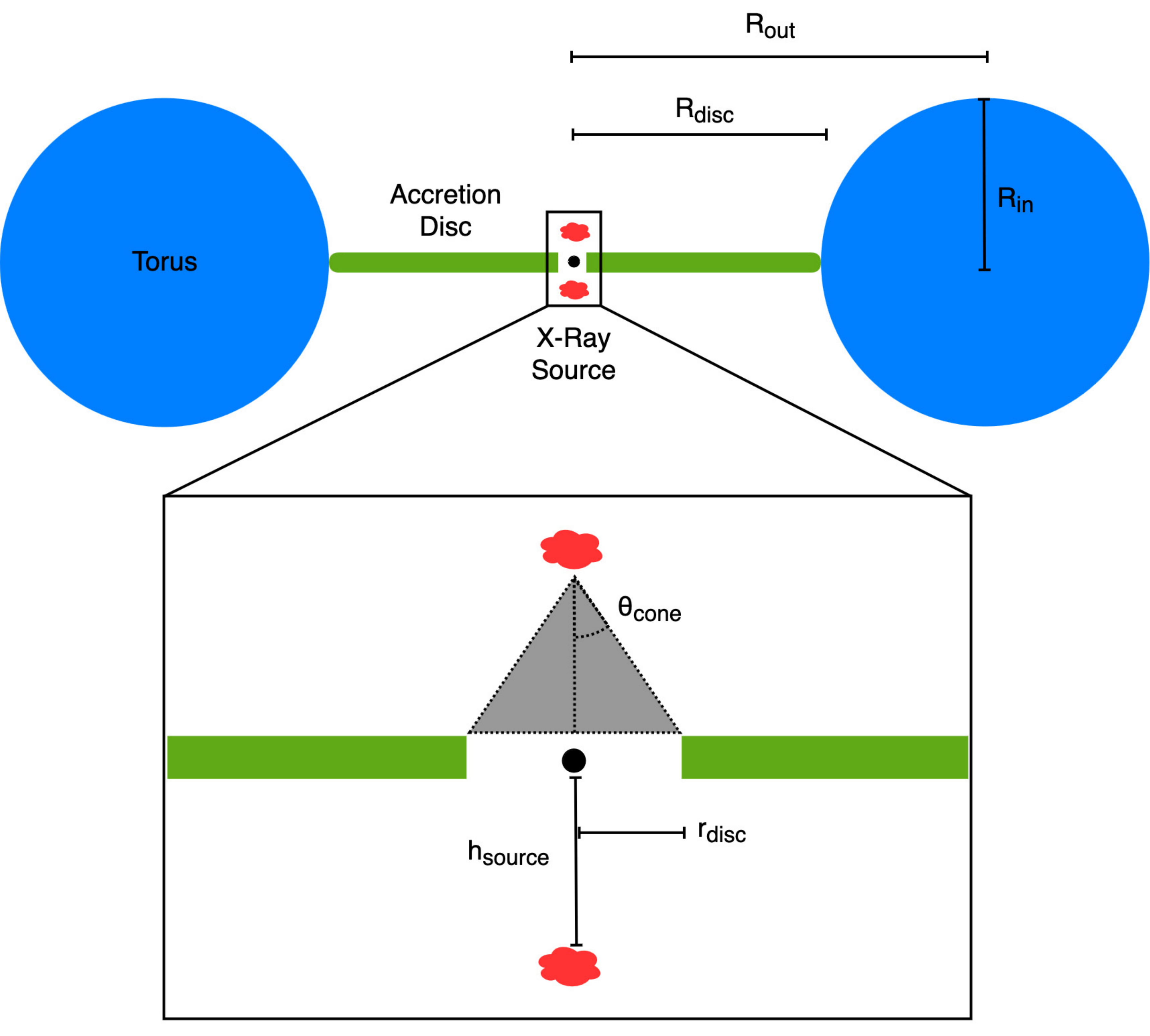}
\caption{Accretion disc and double X-ray source geometry in \textsc{RefleX} (not to scale). The outer radius of the thin annulus representing the AD is set as $R$\textsubscript{disc} = $R$\textsubscript{out} $-$ $R$\textsubscript{in}, with $R$\textsubscript{in} = 1pc, and the inner radius of the annulus as $r$\textsubscript{disc} = 6$r$\textsubscript{g}. We place a double X-ray point source above and below the center of the annulus at $h$\textsubscript{source} = 10$r$\textsubscript{g}. The grey area represents the cone defined by $h$\textsubscript{source} and $r$\textsubscript{disc}, inside which we do not allow the sources to emit as in reality emission through the centre of the annulus would be blocked by the SMBH. The half-opening angle of the cone, $\theta$\textsubscript{cone} = $\arctan(r$\textsubscript{disc} / $h$\textsubscript{source}), is denoted by the dashed lines.
\label{fig:accretiondisc}}
\end{figure}

\subsubsection{Accretion disc}
To make our model more realistic, we add to the previous model a flat, optically thick, dust-free annulus to simulate the presence of an accretion disc (AD). The outer radius of the annulus is set as $R$\textsubscript{disc} $\equiv$ $r$\textsubscript{dust} = $R$\textsubscript{out} $-$ $R$\textsubscript{in}, so that the torus is its physical continuation (Fig. \ref{fig:accretiondisc}). This model (hereafter LD+AD) cannot be scaleless because the height of the X-ray source is not related to other distances, such as the inner edge of the torus. Instead, we fix $R$\textsubscript{in} = 1pc \citep[e.g.,][]{2017NatAs...1..679R} and the inner radius of the annulus to $r$\textsubscript{disc} = 6$r$\textsubscript{g} \citep[e.g.,][]{2011MNRAS.414.1269W}, where $r$\textsubscript{g} = 2\,G\,$M$\textsubscript{BH}\,/\,c$^2$ is the Schwarzschild gravitational radius and assuming a typical SMBH mass of $M$\textsubscript{BH} = $10^7$ M\textsubscript{\(\odot\)} \citep[e.g.,][]{2017ApJ...850...74K}. We set H and He to be fully ionised, although since the disc is completely optically thick, this choice has a negligible effect on the simulated spectra.

For this model, we place a double X-ray point source located above and below the centre of the annulus at $h$\textsubscript{source} = 10$r$\textsubscript{g}, following the lamp-post geometry \citep{1996MNRAS.282L..53M, 2004MNRAS.349.1435M}. For the purposes of this study, we adopt the lamp-post geometry as a widely used and longstanding approximation of the X-ray source despite its known limitations \citep[e.g.,][]{2023MNRAS.525.4735T}, since the geometry of the compact corona has negligible effects on the properties of the reflected radiation. To prevent the escape of emission through the centre of the annulus, which in reality would be blocked by the SMBH, we force the emission of the sources to remain outside the cones defined by $h$\textsubscript{source} and $r$\textsubscript{disc} (as shown in Fig.\ref{fig:accretiondisc}), and we correct the intrinsic luminosity of each object in our population by the factor (1 $-$ cos\,$\theta$\textsubscript{cone}) / 2, where $\theta$\textsubscript{cone} = $\arctan(r$\textsubscript{disc} / $h$\textsubscript{source}) is the half-opening angle of the cones.

\subsection{Synthetic population} \label{subsec:simulations}
After defining the models, we create a large sample of AGN for which we simulate their X-ray emission and determine their absorption properties. For each object in our synthetic population we draw: the redshift and intrinsic 2--10\,keV X-ray luminosity of the X-ray source by sampling the XLF derived in U03, the observation angle, $\theta_\mathrm{obs}$, from a cosine distribution between $0^{\circ}$ and $90^{\circ}$, and the equatorial column density, \NHeq{}, of the torus (Fig. \ref{fig:torus}) from a lognormal distribution, log(\NHeq{}/\,cm$^{-2}) \sim \mathcal{N}(\mu, \sigma$), where $\mu$ and $\sigma$ are the mean and the standard deviation of the distribution respectively, and free parameters in all our models.

We calculate the line-of-sight column density of the homogeneous torus, \NH{}, at a certain observation angle, $\theta_\mathrm{obs}$, as shown in \cite{2017A&A...607A..31P}:
\begin{equation}
N_\mathrm{H} = \left\{
         \begin{array}{ l r }
          N_\mathrm{H,eq} \left(1 - \left(\frac{R_\mathrm{out}}{R_\mathrm{in}}\right)^2 \cos^2\theta_\mathrm{obs}\right)^{1/2},& \cos\theta_\mathrm{obs} < \frac{R_\mathrm{in}}{R_\mathrm{out}} \\
          0 \textrm{~cm}^{-2},& \cos\theta_\mathrm{obs} > \frac{R_\mathrm{in}}{R_\mathrm{out}}
         \end{array}
        \right.
\end{equation}.

We perform numerical simulations using \textsc{RefleX} to create a grid of models of AGN X-ray spectra in the energy range 2--100\,keV, with log(\NHeq{}/\,cm$^{-2}$) between 21 and 26 with a step of 0.1 and log($R$\textsubscript{in}/$R$\textsubscript{out}) between $-2$ and 0 with a step of 0.05. We then interpolate from the grid to obtain the spectrum of any AGN in our population. For practical reasons, the grid is computed at $z$ = 0.0001, and the final spectra are obtained by redshifting the ones interpolated from the grid and rescaling the flux according to the luminosity distance.

To represent the diverse populations from which each observable is derived, we create subsamples of our synthetic population. For the reproduction of the CXB, we draw objects with $L\mathrm{_X^{int}} \geq 10^{41}$\,erg\,s$^{-1}$ and $z \leq 3$, as non-blazar AGN beyond this redshift are expected to contribute only $\sim1\%$ \citep[Figs. 14 and 15 in][]{2014ApJ...786..104U}. For the \NH{} distribution from R17, we sample up to $z = 0.3$, since BAT is only sensitive to objects in the local Universe. Drawing objects at random from the XLF would result in large statistical fluctuations at low redshift, where the volume is small, and at high luminosity, where objects are rare. To avoid these stochastic effects and to remove the effect of cosmic variance in the simulations, we divide the redshift ranges considered for each observable in 10 redshift bins. We then draw in each redshift bin the same large number of objects, $N$\textsubscript{sample}. As a different number of objects, $N$\textsubscript{XLF}, is expected when integrating the log(\NH{}/\,cm$^{-2}) < 24$ XLF, each object is assigned a weight $N$\textsubscript{XLF}\,/\,($N$\textsubscript{sample}$-N$\textsubscript{CT}), where $N$\textsubscript{CT} is the number of CT objects in our population within each redshift bin.

\begin{figure}[t]
\centering
\includegraphics[width=\hsize]{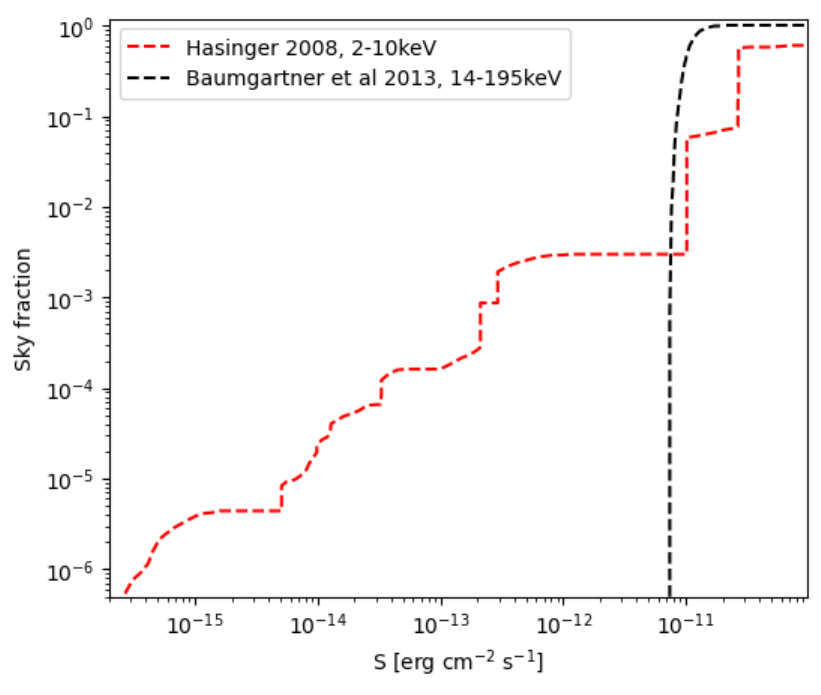}
\caption{Sky fraction as a function of 14--195\,keV and 2--10\,keV flux limits representing the sensitivity curves of the \textit{Swift}/BAT 70-month survey used in R17 (black line) and the multiple surveys used in H08 (red line), respectively.
\label{fig:sensitivity}}
\end{figure}

To correct our synthetic sample for the effect of absorption and compare our results to the observational constraints from R17 and H08, we use the sensitivity curves of the \textit{Swift}/BAT 70-month catalogue \citep{2013ApJS..207...19B} and of the ten samples combined in H08, respectively. We express the sensitivity curves as the sky fraction as a function of 14--195\,keV and 2--10\,keV flux limit for each study, which we show in Fig. \ref{fig:sensitivity}. Using \textsc{RefleX}, we obtain the fluxes of each simulated source in the respective energy bands and apply a weight determined by the probability of such object to be detected according to the sky fraction. This effectively applies the selection functions of these studies on our simulated population.

\section{Results} \label{sec:results}
Following the creation of a synthetic AGN population and the simulation of their X-ray spectra, we determine the parameter posterior distributions of our models based on three data sets: (a) the CXB, (b) the three observed absorption properties, namely the fraction of \NH{} in bins of log(\NH{}) and the fraction of absorbed AGN as a function of observed 2--10\,keV luminosity, and as a function of redshift, and (c) all four constraints combined. 

\subsection{Simulation-based inference} \label{SBI}
We determine model parameters using the simulation-based inference (SBI) approach, as implemented in the Python package sbi \citep{tejero-cantero2020sbi}. SBI maps the relationship between simulated data and model parameters by constructing their joint density in the data and parameter space using a deep neural network. This approach is completely likelihood-free. We add noise to our practically noiseless simulated data to match the uncertainties in the observed data. For the CXB, we add Gaussian noise in each bin using the variance observed in the data. For the absorption properties, the noise is applied in each bin by drawing from a binomial distribution, with the success probability given by the model and the number of trials given by the number of "observed" objects.

We use the Sequential Neural Posterior Estimation (SNPE) \citep{2019arXiv190507488G}, which allows us to directly sample the posterior distribution of our parameters given the data. In SNPE, the density is initially learned across the parameter priors (see Table \ref{tab:models}) with a relatively small number of training samples. We then construct the covariance matrix of the posteriors, which we use to provide a proposal prior in the form of a multivariate normal distribution for a second iteration of SNPE. This process accelerates the convergence of the deep neural network training. Depending on the number of free parameters in each model (three or four), we train the neural estimator using 2000 and 5000 simulations, respectively, to construct the proposal prior; then we retrain with 5000 and 10000 simulations, respectively, using the proposal prior. We find that these sizes for the training sets provide stable posterior estimates.

While model comparison can be performed in principle with any parameter inference 
algorithm that generates samples from the posterior \citep[e.g.,][]{2021arXiv211112720M}, we adopt here a simple metric to compare the performance of the different models; namely the $\chi^2$ statistic of the median models drawn from the posteriors. Indeed, we do not expect that the simple models we explore are able to fully represent the data, so that a statistically rigorous model comparison is not warranted.

\subsection{CXB} \label{subsec:simpletorusresults}

\begin{table*}[t]
	\begin{center}
 	\caption{Median values of the parameter posterior distributions and reduced $\chi^2$ values for each model applied on the three  observational data sets.} \label{table:simpletorusfits}
    \setlength\extrarowheight{4pt}
	\begin{tabular}{l c c c c c c c}
    \hline\hline
		\rule{0pt}{1em}Model & Constraints & $\mu$ (log/\,cm$^{-2}$) & $\sigma$ (log/\,cm$^{-2}$) & $R_0$\tablefootmark{a} & $\alpha$ & $R_1$\tablefootmark{a} & $\chi^2$/dof \\
		\hline
        Simple torus & CXB & 24.43$\mathrm{_{-0.01}^{+0.01}}$ & 0.31$\mathrm{_{-0.02}^{+0.01}}$ & - & - & 0.24$\mathrm{_{-0.01}^{+0.01}}$ & 226/84 \\
		Simple torus & Absorption & 23.27$\mathrm{_{-0.01}^{+0.01}}$ & 0.10$\mathrm{_{-0.03}^{+0.01}}$ & - & - & 0.98$\mathrm{_{-0.05}^{+0.03}}$ & 742/22 \\
        LD torus & Absorption & 23.77$\mathrm{_{-0.07}^{+0.08}}$ & 1.06$\mathrm{_{-0.09}^{+0.08}}$ & 0.37$\mathrm{_{-0.01}^{+0.02}}$ & 0.41$\mathrm{_{-0.02}^{+0.01}}$ & - & 122/21 \\
		LD torus & Combined & 24.24$\mathrm{_{-0.02}^{+0.03}}$ & 0.68$\mathrm{_{-0.01}^{+0.02}}$ & 0.21$\mathrm{_{-0.01}^{+0.01}}$ & 0.22$\mathrm{_{-0.01}^{+0.01}}$ & - & 1231/109 \\
		LD+AD & Combined & 23.56$\mathrm{_{-0.04}^{+0.05}}$ & 0.98$\mathrm{_{-0.11}^{+0.09}}$ & 0.93$\mathrm{_{-0.03}^{+0.04}}$ & 0.24$\mathrm{_{-0.01}^{+0.01}}$ & - & 804/109 \\
        LD+AD+$E$\textsubscript{C}\tablefootmark{b} & Combined & 23.56$\mathrm{_{-0.04}^{+0.05}}$ & 0.98$\mathrm{_{-0.11}^{+0.09}}$ & 0.93$\mathrm{_{-0.03}^{+0.04}}$ & 0.24$\mathrm{_{-0.01}^{+0.01}}$ & - & 540/109 \\[2pt]
	\hline
    \end{tabular}
    \tablefoot{\tablefoottext{a}{Arbitrary units for models without AD and measured in pc for the AD model.}\\
    \tablefoottext{b}{The same parameter set as before but with a fixed high energy cut-off of $E$\textsubscript{C} = 300\,keV (see \ref{subsec:recedingdiscresults} for more information).}}
	\end{center}
\end{table*}

\begin{figure}[t]
\centering
\subfloat{\includegraphics[width = 1.\hsize]{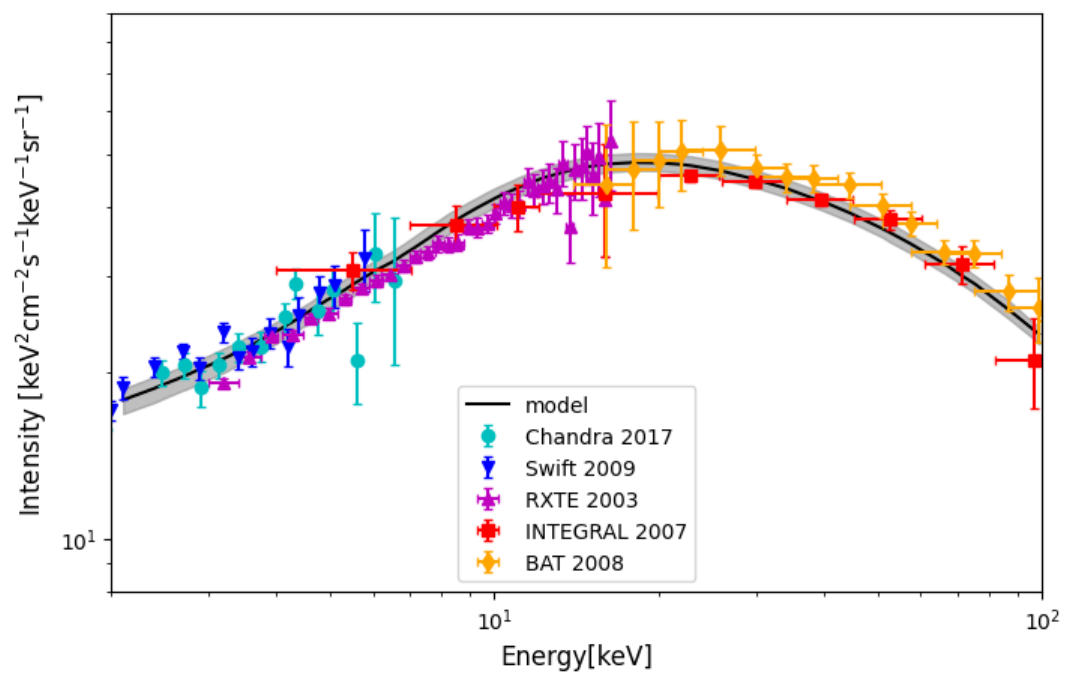}}
\caption{Median model (black line) for the CXB in the range 2--100\,keV using the simple torus applied only on the CXB data set. The grey region represents the 68\% confidence interval of the model based on the posterior distribution of the parameters.
\label{fig:cxbresults}}
\end{figure}

We first test the ability of our models to reproduce the CXB data set. As shown in Fig. \ref{fig:cxbresults}, the simple torus model is sufficient to match the CXB, proving that the CXB can be reproduced by even the simplest AGN unification models and, consequently, does not possess a strong constraining power. In Table \ref{table:simpletorusfits}, we summarise the median values of the parameter posteriors and the corresponding reduced $\chi^2$ values. All posterior distributions are presented in Appendix \ref{app:sbiposteriors}. In this AGN population, if observed edge-on, almost 92\% of the AGN would be CT, since the lognormal distribution of \NHeq{} is centred around log(\NHeq{}/\,cm$^{-2}$) = 24.4 with a scatter of only a factor of $\sim$ 1.4.

The LD torus model can reproduce the CXB as well with model parameters close to those of the simple torus model and the posterior distribution of the slope in Eq. (\ref{eq:RoutLx}) showing a median value near 0.

\subsection{Absorption properties} \label{subsec:recedingresults}

\begin{figure*}[t]
\centering
\subfloat{\includegraphics[width = 0.314\hsize]{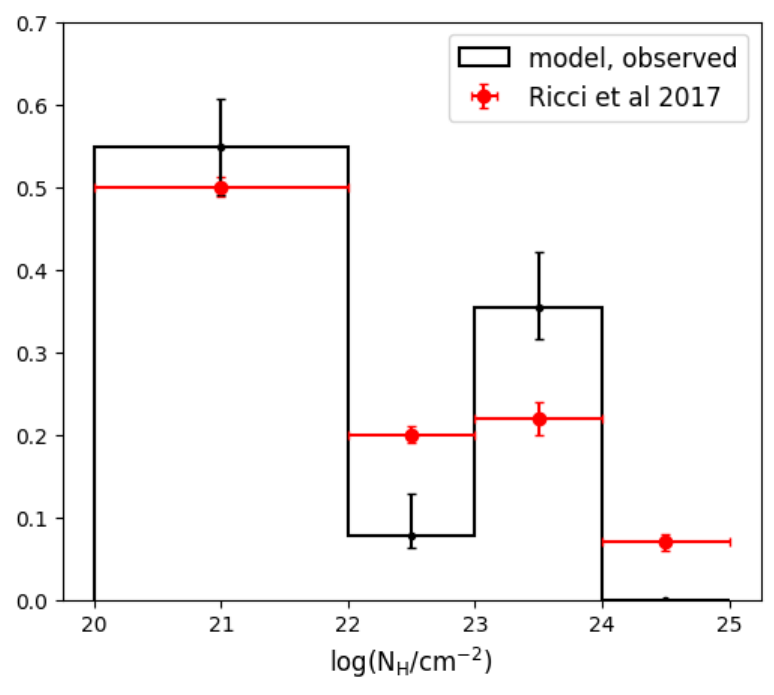}} 
\subfloat{\includegraphics[width = 0.33\hsize]{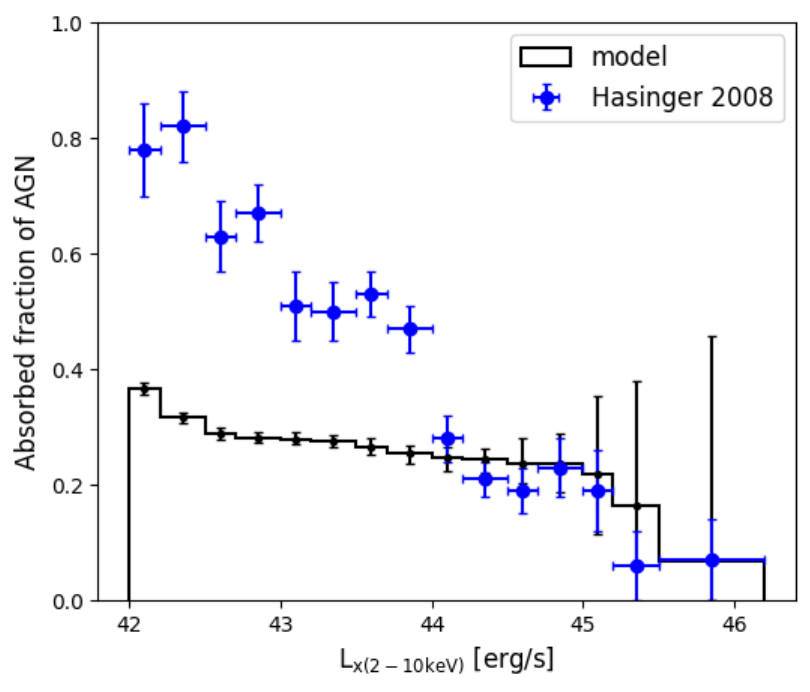}}
\subfloat{\includegraphics[width = 0.332\hsize]{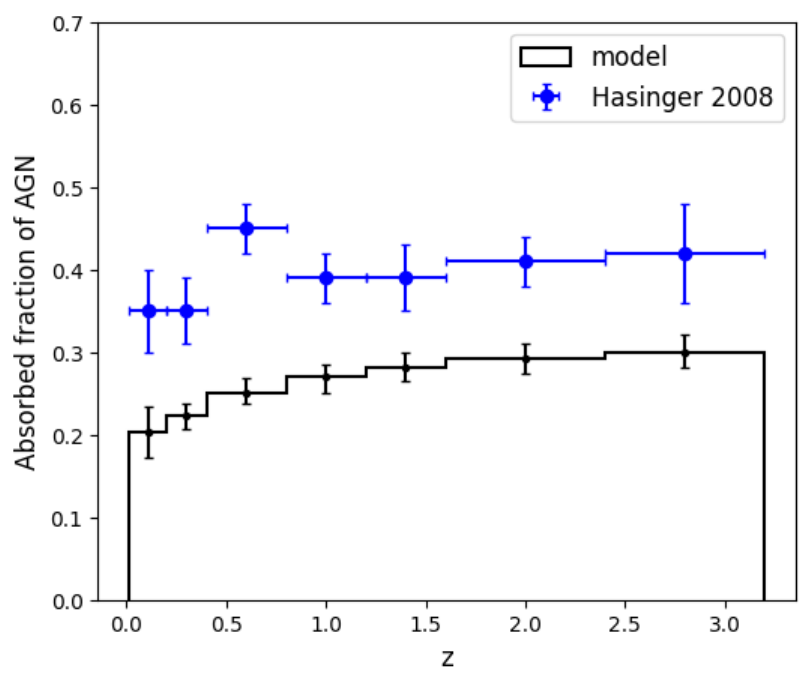}}\\
\subfloat{\includegraphics[width = 0.314\hsize]{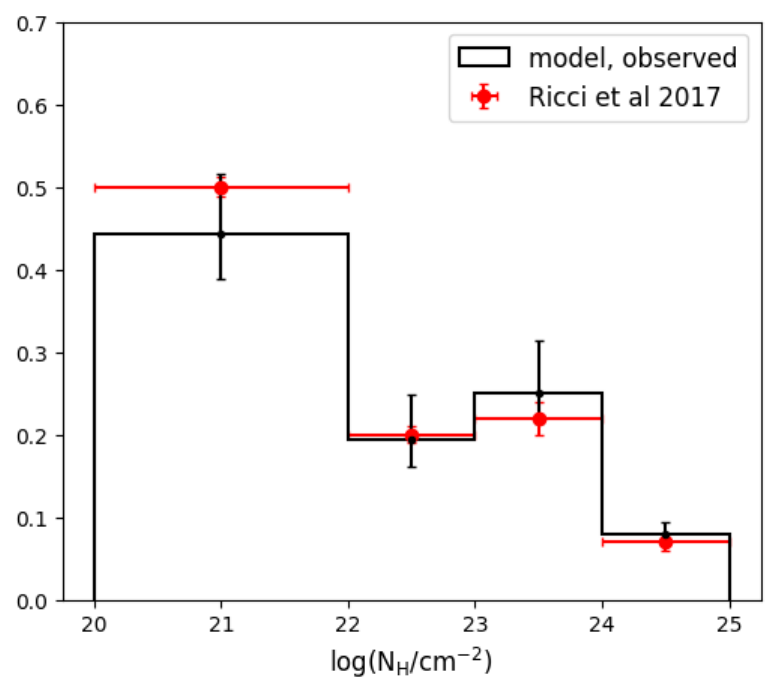}}
\subfloat{\includegraphics[width = 0.33\hsize]{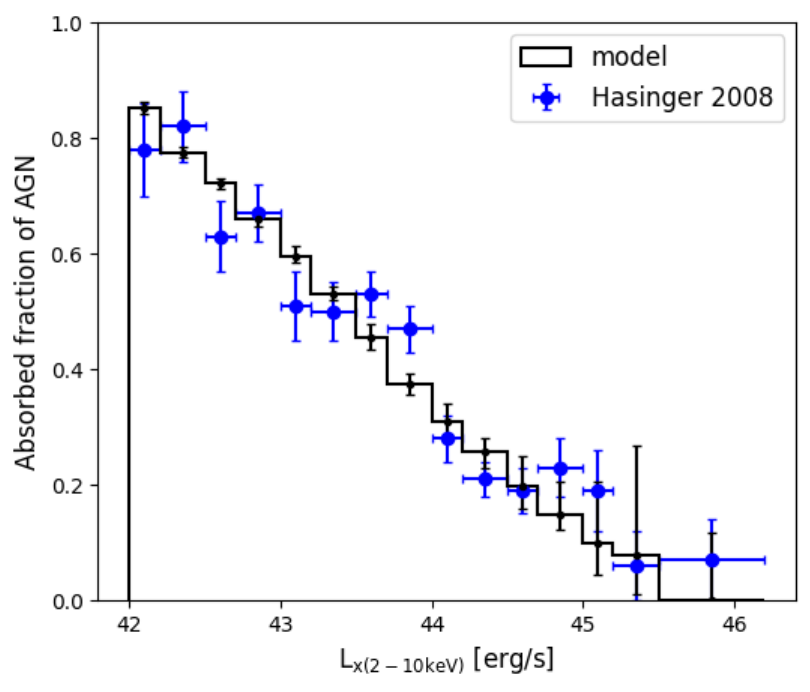}}
\subfloat{\includegraphics[width = 0.332\hsize]{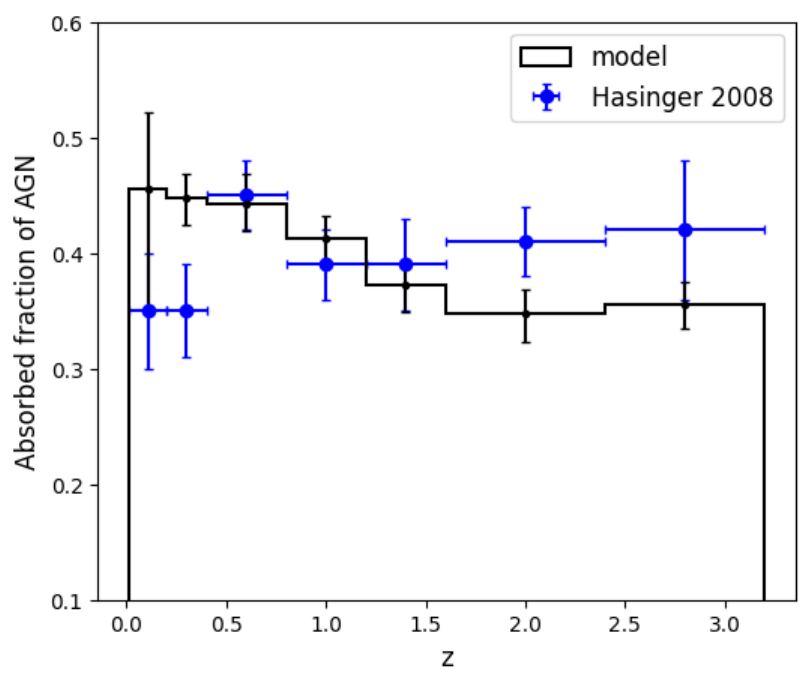}}\\
\caption{Median models (black lines) for the three observed absorption properties using the simple torus model (top row) and the LD torus (bottom row) applied only on the three absorption data sets: \textit{left:} the fraction of \NH{} in bins of log(\NH{}), \textit{middle:} the absorbed fraction of AGN as a function of observed L\textsubscript{X(2--10\,keV)}, and \textit{right:} as a function of redshift. The error bars on the models represent the 68\% confidence interval of the models based on the posterior distribution of the parameters.
\label{fig:absresults}}
\end{figure*}

We next apply our models on the three absorption data sets. The limitations of the simple torus model become evident when applied to the absorption properties, even when the CXB constraint is ignored. In this case, the model fails to reproduce the observed anti-correlation between the absorbed fraction and luminosity (top row of Fig. \ref{fig:absresults}).

In contrast, the LD torus model reproduces very well the three constraints (bottom row of Fig. \ref{fig:absresults}). In this case, the posterior distribution of the slope in Eq. (\ref{eq:RoutLx}) yields a median value of $\alpha$ = 0.41$\mathrm{_{-0.02}^{+0.01}}$, close to, but still different from, the predictions of the receding torus model. Furthermore, matching the absorption properties requires the lognormal distribution of \NHeq{} to be more than three times broader and centred 0.66 dex lower compared to the optimal match on the CXB. This, combined with the strong luminosity dependence of the covering factor, results in a distinctly different AGN population than the one that best reproduces the CXB. Specifically, for the CXB, approximately 55\% of the synthetic population needs to be CT and only $\sim13\%$ Compton-thin, whereas for the absorption properties, the fractions shift to around 33\% CT and 51\% Compton-thin.

\subsection{Combined constraints} \label{subsec:recedingdiscresults}

\begin{figure*}[t]
\centering
\subfloat{\includegraphics[width = 0.5\hsize]{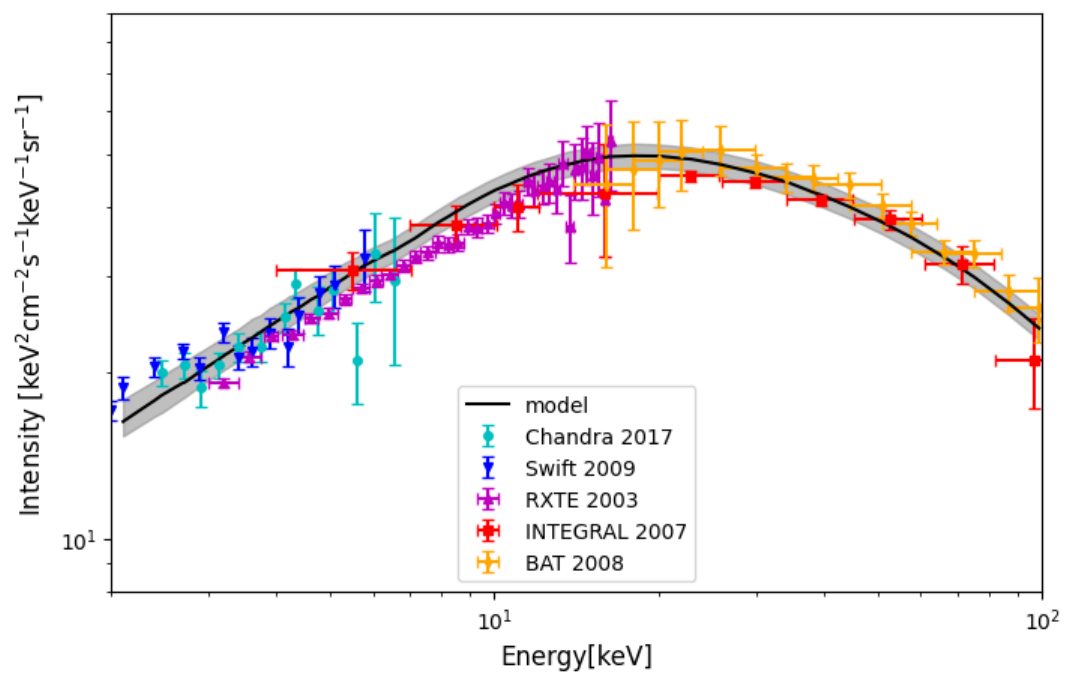}} \\
\subfloat{\includegraphics[width = 0.314\hsize]{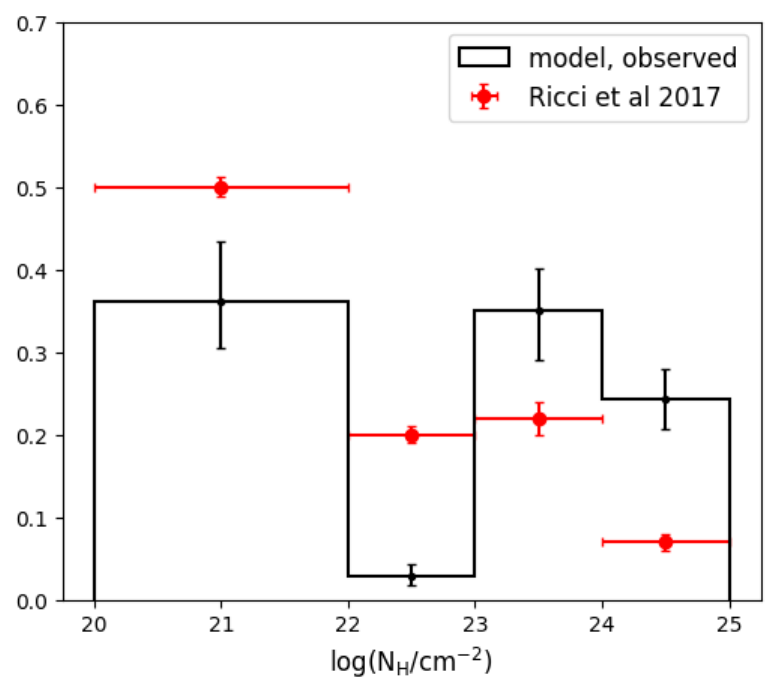}}
\subfloat{\includegraphics[width = 0.33\hsize]{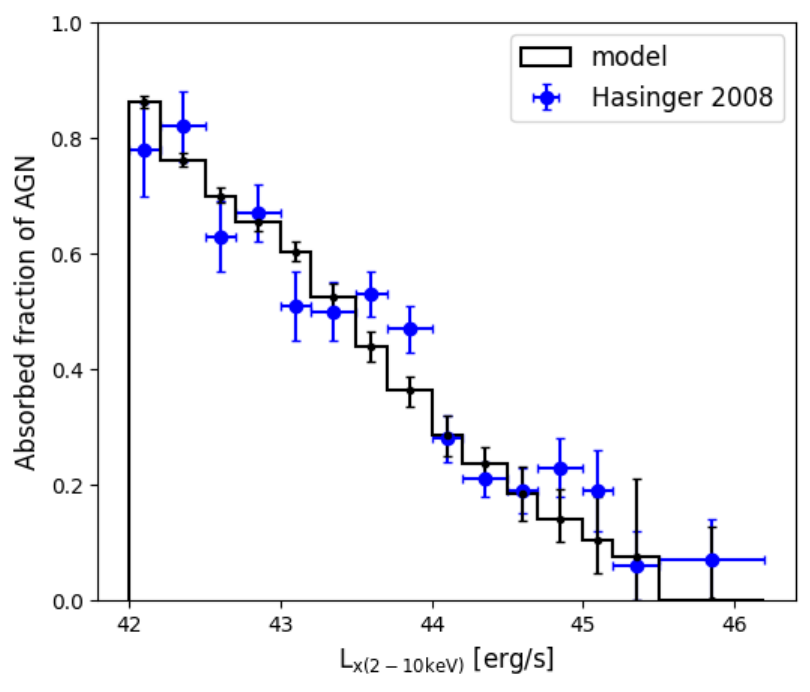}}
\subfloat{\includegraphics[width = 0.332\hsize]{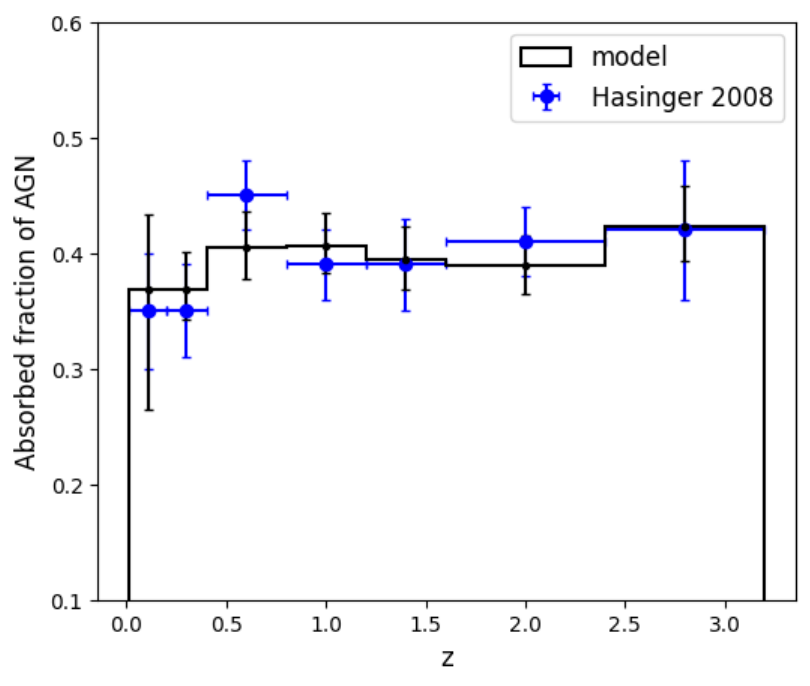}}
\caption{Median models (black lines) using the LD torus model applied on the four combined constraints: \textit{top:} the CXB, \textit{bottom left:} the observed fraction of \NH{} in bins of log(\NH{}), \textit{bottom middle:} the absorbed fraction of AGN as a function of observed L\textsubscript{X(2--10\,keV)}, and \textit{bottom right:} as a function of redshift. The grey region and error bars represent the 68\% confidence interval of the model based on the posterior distribution of the parameters.
\label{fig:reccombinedresults}}
\end{figure*}

\begin{figure*}[t]
\centering
\subfloat{\includegraphics[width = 0.5\hsize]{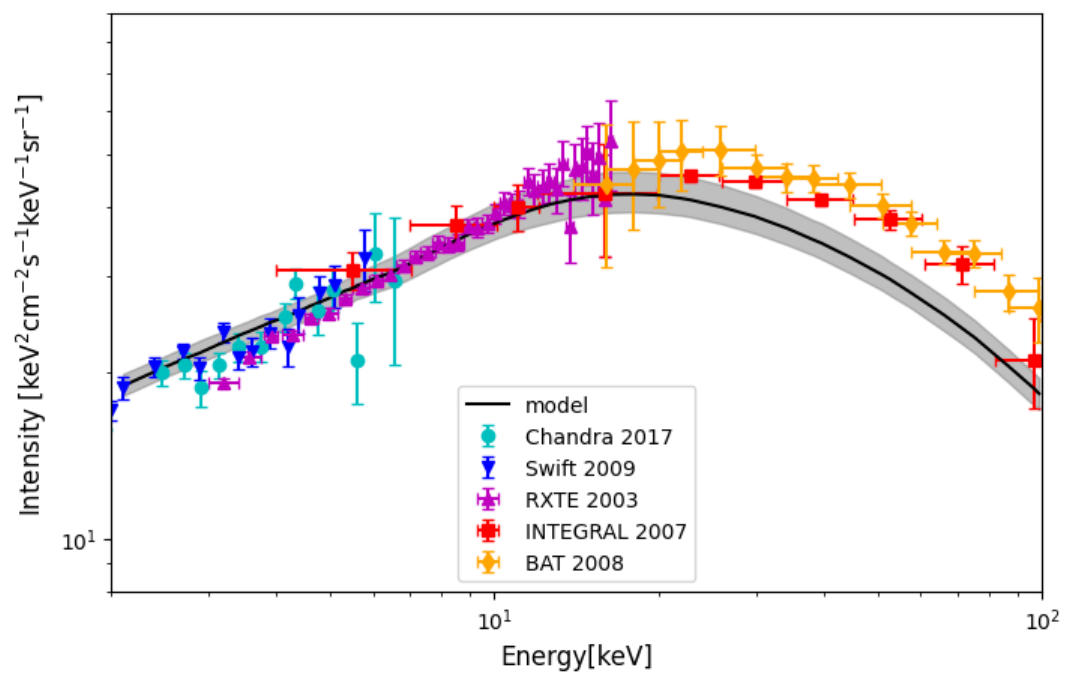}}
\subfloat{\includegraphics[width = 0.5\hsize]{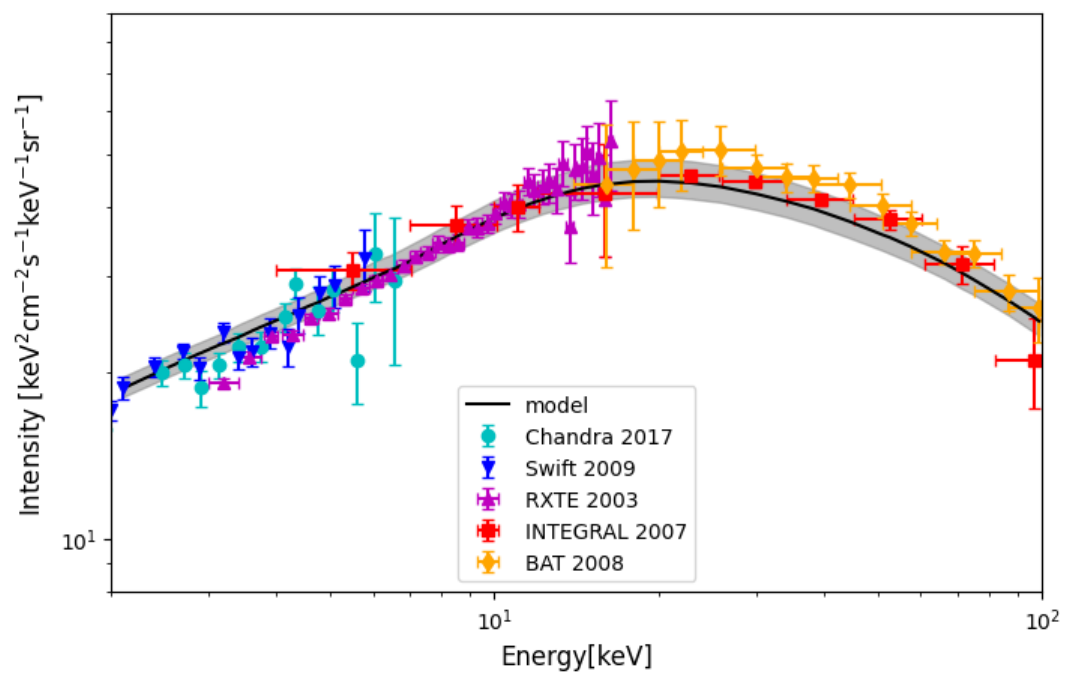}} \\
\subfloat{\includegraphics[width = 0.314\hsize]{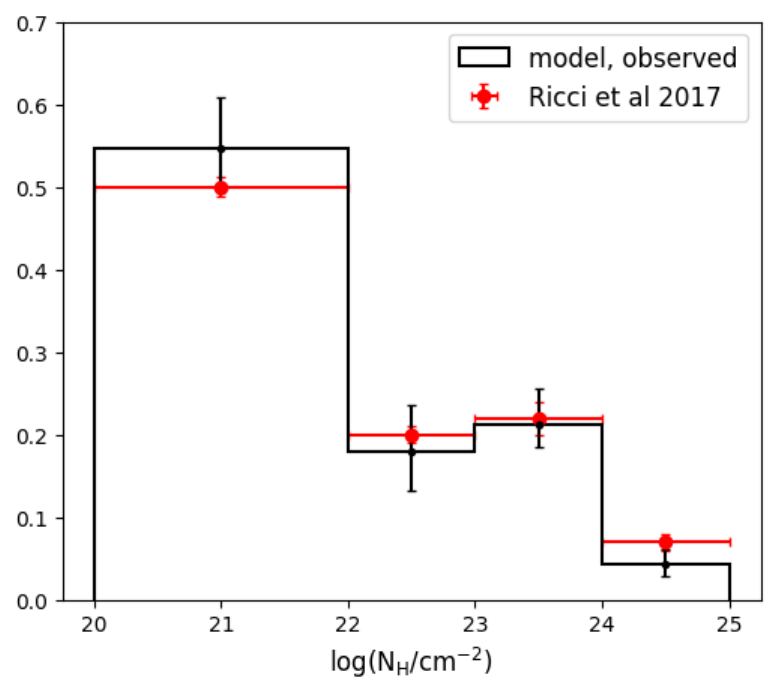}}
\subfloat{\includegraphics[width = 0.33\hsize]{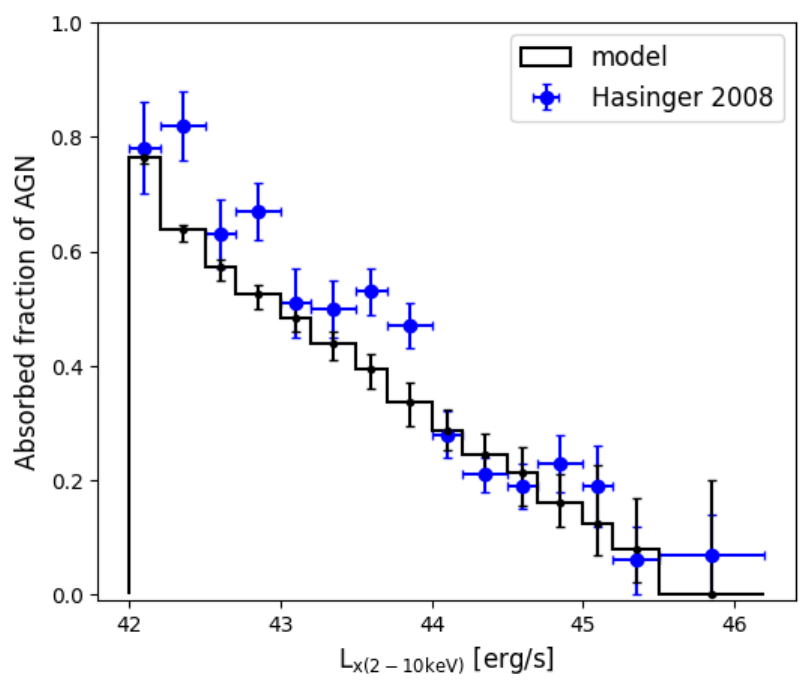}}
\subfloat{\includegraphics[width = 0.332\hsize]{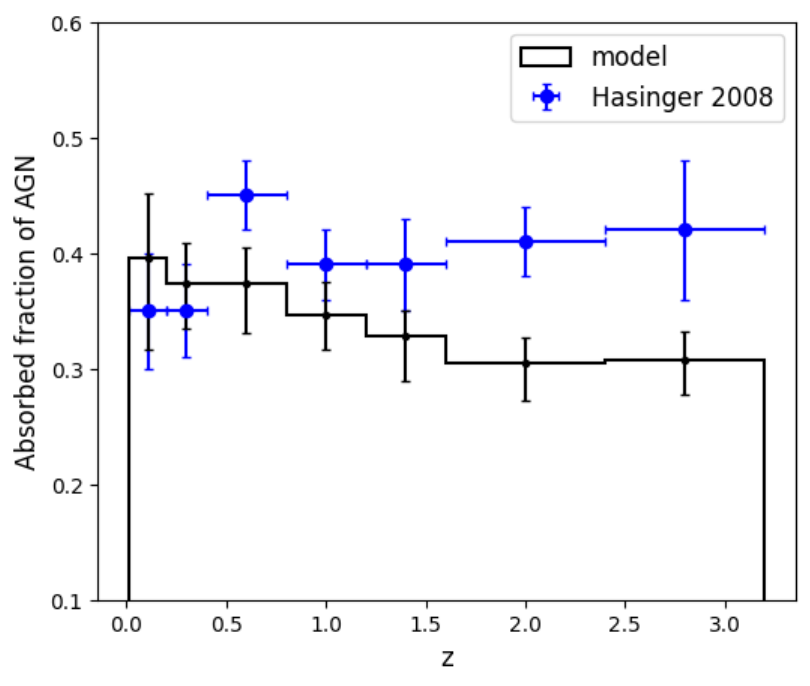}}
\caption{Same as Fig. \ref{fig:reccombinedresults} but using the LD+AD model. \textit{Top left:} the CXB with fixed high-energy cutoff $E$\textsubscript{C} = 200\,keV, and \textit{top right:} with $E$\textsubscript{C} = 300\,keV (see \ref{subsec:recedingdiscresults} for more information).
\label{fig:recdisccombinedresults}}
\end{figure*}

Finally, after demonstrating the distinct differences between the two AGN populations that best match the CXB and the absorption properties separately, we seek one population that satisfies all four constraints simultaneously. Given that the simple torus model clearly fails to reproduce the absorption properties, we do not apply it on the combined data sets, focusing instead on the LD torus model. For simplicity, we calculate the combined $\chi^2$ without applying any particular weight on the data sets, despite the heteroscedasticity of the combined set. In Fig. \ref{fig:reccombinedresults}, we show the CXB and the absorption properties reproduced by the LD torus. While most constraints are adequately matched, the model completely fails to explain the shape of the \NH{} fraction in bins of log(\NH{}), underestimating the number of less obscured objects and overestimating that of heavily obscured objects. The failure of the LD model to reproduce our combined constraints highlights the need for a more complex model to better capture the observed properties of the AGN population.

Building on the limitations of the LD torus model, we now turn to the LD+AD model to assess whether it can provide a better match to the combined observational constraints. While this model provides the best simultaneous match to all four constraints (shown in Fig. \ref{fig:recdisccombinedresults}), it significantly underpredicts the CXB at high energies and predicts a $\sim$25\% lower absorbed AGN fraction at high redshifts. The posterior distribution of the slope in Eq. (\ref{eq:RoutLx}) yields a median value of $\alpha$ = 0.24$\mathrm{_{-0.01}^{+0.01}}$, substantially deviating from the predictions of the receding torus model but in close agreement with X-ray observations \citep[e.g.,][]{2011ApJ...728...58B, 2015MNRAS.454.1202S}. The lognormal distribution of \NHeq{} of this population aligns very closely with that of the best match on the absorption properties, rather than the CXB, further demonstrating that the CXB does not carry enough constraining power. Even then, the posterior distributions of the normalisation, $R$\textsubscript{0}, and the slope, $\alpha$, in Eq. (\ref{eq:RoutLx}) have medians that are more than twice as high and nearly twice as flat, respectively, compared to the best match on the absorption properties. The combined constraints result in the lowest fraction of CT AGN ($\sim$21\%) and a Compton-thin fraction of $\sim$48\%, leaving an unobscured fraction of $\sim$31\%.

To address the underpredicted CXB at high energies, we increase the fixed high-energy cutoff of the X-ray source from $E$\textsubscript{C} = 200\,keV to $E$\textsubscript{C} = 300\,keV, as low accretion AGN are more common and tend to have higher $E$\textsubscript{C} \citep[R17,][]{2018MNRAS.480.1819R,2020ApJ...905...41B}. Since variations in $E_{\text{C}}$ affect the high-energy region of the AGN spectra while having minimal impact on the lower energy range, we find that redetermining model parameters does not result in significant changes to the parameter posteriors. Using the same parameter posteriors, we find that the match on the CXB improves considerably (top right in Fig. \ref{fig:recdisccombinedresults}), while the absorption properties remain consistent.

\section{Discussion} \label{sec:discussion}
Our results demonstrate that the CXB can be reproduced not only by more complex models, but even by the simple unification model of AGN. However, our simple torus model lacks some necessary ingredients to accurately represent the absorption properties. First, the covering factor of the absorber must have a clear luminosity dependence, which cannot be constrained by the CXB alone. Second, the inclusion of the AD component is essential to provide enough reflection given the constraint of the \NH{} distribution. Hence, the LD+AD model is the only realistic model capable of simultaneously matching both the CXB and the absorption data sets. Nevertheless, all models discussed in this work are simplistic and cannot be expected to explain the full phenomenology of AGN in detail.

\subsection{Luminosity dependence of absorbed fraction} \label{subsec:absorptiondependence}

\begin{figure}[t]
\centering
\includegraphics[width=\hsize]{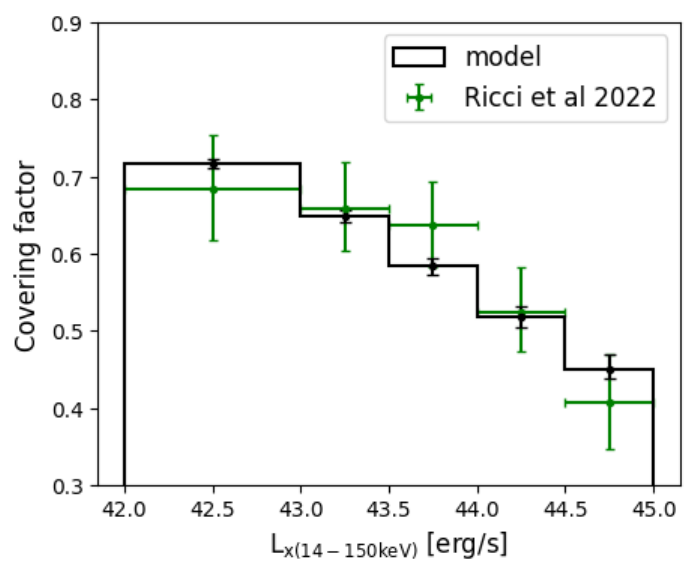}
\caption{Average torus covering factor of our local synthetic population as a function of intrinsic 14--150\,keV luminosity using the LD+AD model (black). The error bars represent the 68\% confidence interval of the model based on the posterior distribution of the parameters. For comparison, the covering factor of \textit{Swift}/BAT detected AGN from \customcite{2022ApJ...938...67R}{green}.
\label{fig:coveringfactor}}
\end{figure}

As shown in Sect. \ref{subsec:recedingdiscresults}, when applying the LD+AD model to all observational constraints simultaneously, we are able to reproduce the dependence of the absorbed fraction on the X-ray luminosity. From parameter determination, we can derive that the inner edge of typical dusty tori, $r$\textsubscript{dust} = $R$\textsubscript{out} $-$ $R$\textsubscript{in}, spans a range of 0.2--2.8\,pc in the L$\mathrm{_X^{int}}$ range 10$^{41} - 10^{46}$\,erg\,s$^{-1}$, consistent with dust sublimation radius sizes found in infrared interferometry studies \citep[e.g.,][]{2014A&A...563A..82T, 2020A&A...635A..92G}. However, we find a shallow median slope in Eq. (\ref{eq:RoutLx}), $\alpha$ = 0.24$\mathrm{_{-0.01}^{+0.01}}$, compatible with observational studies in the X-rays \citep[e.g.,][]{2011ApJ...728...58B, 2015MNRAS.454.1202S}, but in disagreement with the receding torus model. To further investigate, Fig. \ref{fig:coveringfactor} shows the average covering factor of the detected tori in our local AGN population as a function of intrinsic 14-150\,keV luminosity to compare with the \textit{Swift}/BAT results from \cite{2022ApJ...938...67R}. We find that the covering factor decreases by 27\% across the $L\mathrm{_{X(14-150keV)}}$ range of 10$^{42} - 10^{45}$ erg s$^{-1}$, in very good agreement with \cite{2022ApJ...938...67R}, despite our model being constrained by the absorbed fraction from H08, which comprises a population up to z = 3.2 detected in the 2--10\,keV.

Other mechanisms have been proposed to explain the luminosity dependence of the absorbed fraction with a slope flatter than that predicted by the receding torus model. Additional X-ray obscuration could arise from the broad-line region (BLR), which is located within the dust sublimation radius. The presence of absorption in the BLR could increase the fraction of absorbed AGN at high luminosity, thus flattening the relationship between the absorbed fraction and luminosity. Another important mechanism is the clumpy dusty torus model \citep[e.g.,][]{2008ApJ...685..160N}, from which \cite{2007MNRAS.380.1172H} predict that the absorbed fraction decreases as $\propto L^{-0.25}$ due to the interplay between radiation pressure and the clump distribution. It must be pointed out that the recent studies of \cite{2017Natur.549..488R, 2022ApJ...938...67R} have revealed that when the AGN are binned by Eddington ratio, the luminosity dependence of the absorbed fraction completely flattens. Instead, they propose a radiation-regulated unification model, where the absorbed fraction depends on the Eddington ratio, rather than luminosity. 

\begin{figure}[t]
\centering
\subfloat{\includegraphics[width = \hsize]{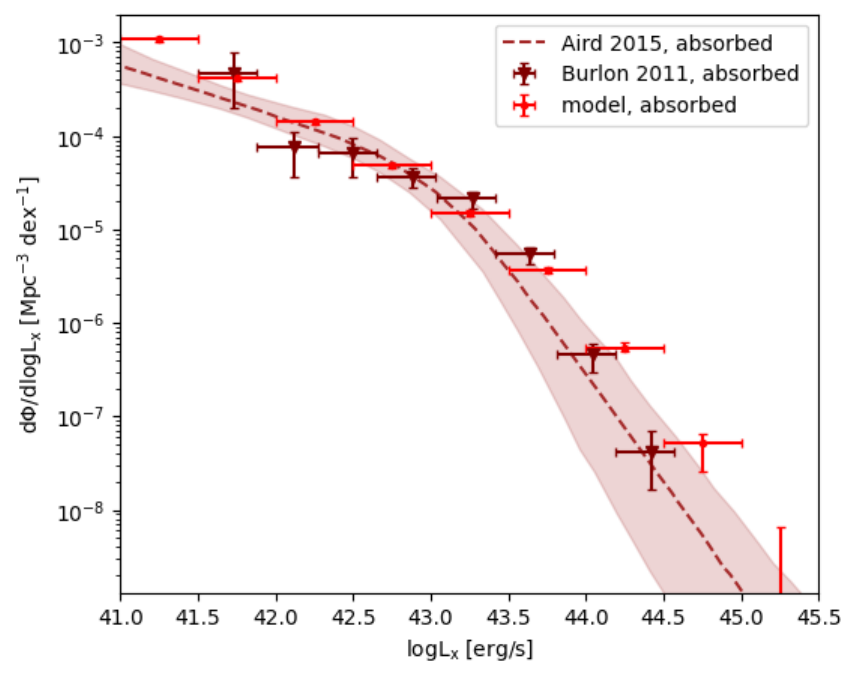}}\\ 
\subfloat{\includegraphics[width = \hsize]{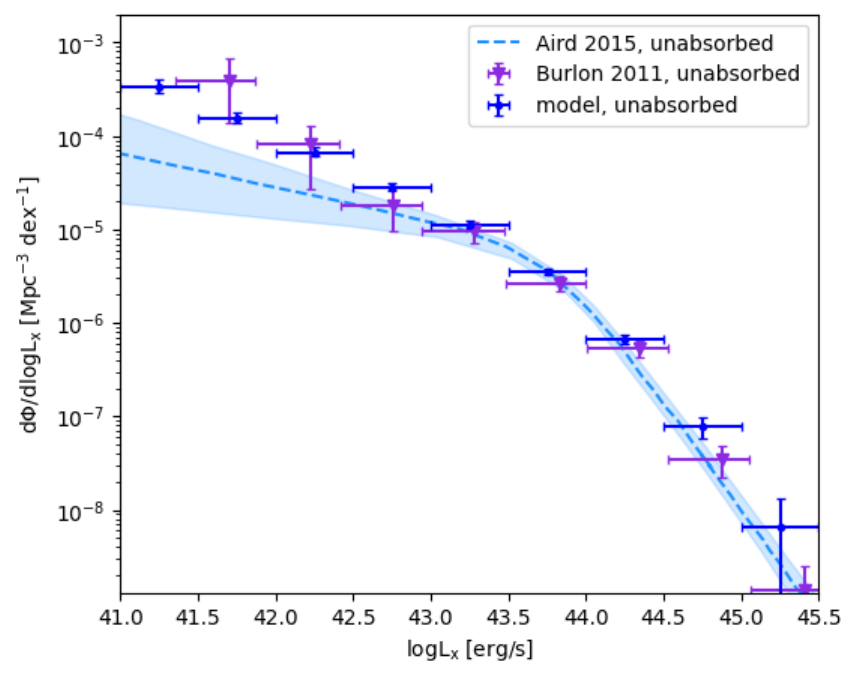}}
\caption{X-ray luminosity function of absorbed (red points) and unabsorbed (blue points) AGN in our population up to z = 0.1, using the LD+AD model with our combined constraints. The error bars represent the 68\% confidence interval of the model based on the posterior distribution of the parameters. \textit{Top:} Comparison of the absorbed XLF with Fig. 13 of \cite{2015MNRAS.451.1892A}, where they extrapolate their model at z=0 (brown dashed line) and present the local observations of \cite{2011ApJ...728...58B} converted to 2--10\,keV luminosities (brown triangles). \textit{Bottom:} Same comparison of the unabsorbed XLF with  \cite{2015MNRAS.451.1892A} (blue dashed line) and \cite{2011ApJ...728...58B} (purple triangles).
\label{fig:XLFtypes}}
\end{figure}

As a direct consequence of the relation between absorption and luminosity, our LD+AD model predicts that absorbed and unabsorbed AGN have different XLFs. In Fig. \ref{fig:XLFtypes}, we present the two XLFs of our population up to z = 0.1. Our model predicts well the increase of absorbed AGN below log($L$\textsubscript{X}) = 43.5, as found by the model of \cite{2015MNRAS.451.1892A} extrapolated to z = 0 (taken from their Fig. 13) and the local observations in \customcite{2011ApJ...728...58B}{converted to 2--10\,keV luminosities}. Our model confirms that absorbed AGN are intrinsically less luminous. Both our absorbed and unabsorbed XLFs agree very well with the results of \cite{2011ApJ...728...58B}, but our unabsorbed XLF deviates from the results of \cite{2015MNRAS.451.1892A} at low luminosities. That may be because the latter is mostly constrained by high-redshift sources.

\subsection{AGN number counts} \label{subsec:counts}

\begin{figure*}[t]
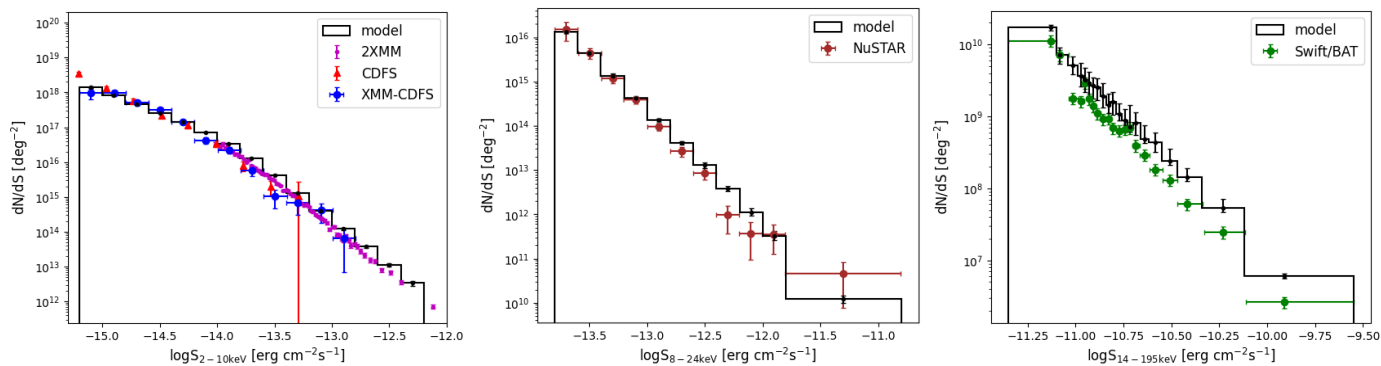

\centering
\subfloat{\includegraphics[width = 0.335\hsize]{Images/XMM-dN.pdf}}
\subfloat{\includegraphics[width = 0.325\hsize]{Images/Nustar-dN.pdf}}
\subfloat{\includegraphics[width = 0.33\hsize]{Images/Swift-dN.pdf}}\\
\caption{Differential number counts as a function of flux produced by the LD+AD model (black) in the \textit{left:} 2--10\,keV, \textit{centre:} 8--24\,keV and \textit{right:} 14--195\,keV energy ranges. The error bars represent the 68\% confidence interval of the model based on the posterior of the parameters. For comparison we present the differential number counts of the 70-month \textit{Swift}/BAT \citep[][green]{2022ApJS..261....9A}, combined NuSTAR \citep[][brown]{2016ApJ...831..185H}, 4\,Ms CDFS \citep[][red]{2012ApJ...752...46L}, XMM-CDFS \citep[][blue]{2013A&A...555A..42R}, and 2XMM \citep[][magenta]{2008A&A...492...51M} surveys.
\label{fig:counts}}
\end{figure*}

In Fig. \ref{fig:counts}, we present AGN differential number counts as a function of flux in the 2--10\,keV, 8--24\,keV and 14--195\,keV energy bands as reproduced by the LD+AD model. The model predicts well the 2--10\,keV number counts from the 4\,Ms CDFS \citep{2012ApJ...752...46L}, XMM-CDFS \citep{2013A&A...555A..42R}, and 2XMM \citep{2008A&A...492...51M} surveys, particularly at fainter fluxes. Similarly in the 8--24keV band, the model can reproduce the number counts from the combined surveys of NuSTAR \citep{2016ApJ...831..185H}.

In contrast, the model consistently overpredicts the 14--195\,keV \textit{Swift}/BAT differential number counts \citep{2022ApJS..261....9A} within 1$\sigma$--2$\sigma$. We note that this is in agreement with several other AGN population synthesis models based on the CXB \citep[e.g.,][]{2007A&A...463...79G,2014ApJ...786..104U,2015MNRAS.451.1892A}, which also produce higher number counts than \textit{Swift}/BAT, as shown in \cite{2019ApJ...871..240A} and \cite{2016ApJ...831..185H}. This tension is still under investigation \citep[e.g.][]{2019A&A...625A.131A, 2023MNRAS.521.1620D}.

\subsection{Reflection vs obscuration} \label{subsec:reflection}

\begin{figure}[t]
\centering
\includegraphics[width=\hsize]{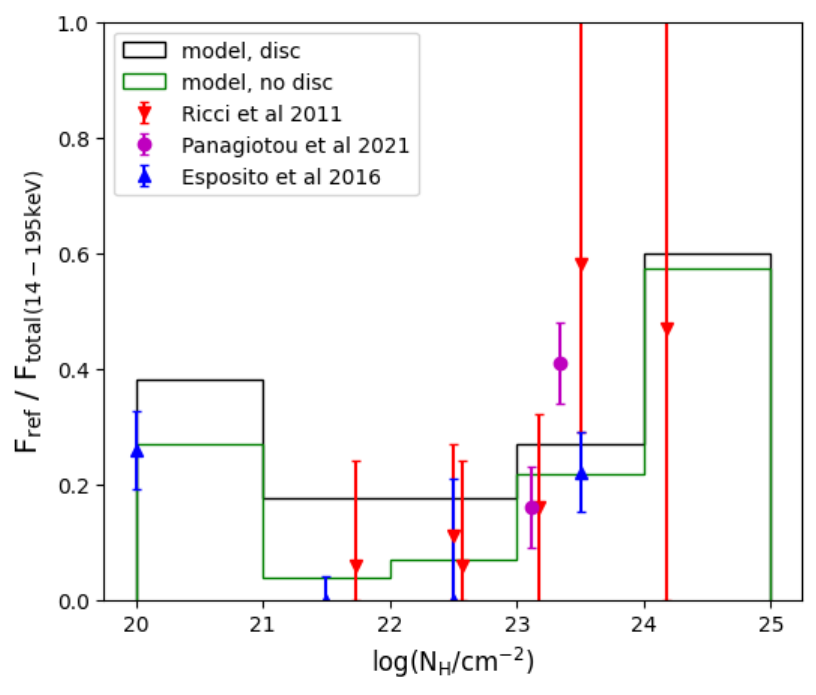}
\caption{Observed correlation between the fraction of reflected emission over total emission, $F$\textsubscript{ref} / $F$\textsubscript{total}, in the energy band 14--195\,keV and the line-of-sight \NH{}, produced with our combined constraints median parameters for the LD+AD model (black line) and the LD torus model (green line). The data are from \cite{2021A&A...653A.162P}, \cite{2016A&A...590A..49E} and \cite{2011A&A...532A.102R} (magenta, blue and red points, respectively), after replacing the reflection parameter of the \textit{pexrav} model in XSPEC with the fraction $F$\textsubscript{ref} / $F$\textsubscript{total} in the same energy band.
\label{fig:ref-abs}}
\end{figure}

In Fig. \ref{fig:ref-abs}, we compare the predicted reflection of our models to the observed trend between reflection and line-of-sight \NH{} reported in \cite{2021A&A...653A.162P}, \cite{2011A&A...532A.102R}, and \cite{2016A&A...590A..49E}. We compute the fraction of reflected to total emission ($F$\textsubscript{ref} / $F$\textsubscript{total}) in the 14--195\,keV band using \textsc{RefleX} for our LD and LD+AD models. Although these previous studies suggest an increase of reflection with obscuration, large uncertainties and unknown selection effects make this correlation debated and not statistically conclusive. We therefore use the data for qualitative comparison only, by replacing the reflection parameter of the \textit{pexrav} model in XSPEC \citep{1996ASPC..101...17A} with $F$\textsubscript{ref} / $F$\textsubscript{total} in the same energy band.

Both models broadly follow the trend of increasing reflection with obscuration, while they also predict the increase of reflection for unobscured sources [log(\NH{}/\,cm$^{-2}$) < 21], which is suggested in \cite{2016A&A...590A..49E} and \cite{2019A&A...626A..40P}. The key difference between our models arises at log(\NH{}/\,cm$^{-2}$) < 23, where the LD+AD model consistently provides optically thick material necessary to produce reflection. In comparison, the LD model lacks such material in this \NH{} range, resulting in nearly half the reflection predicted by the LD+AD model.

While \cite{2011A&A...532A.102R} and \cite{2021A&A...653A.162P} attribute enhanced reflection in heavily obscured sources to a clumpy dusty torus, we demonstrate that such a trend can also emerge from our homogeneous torus models. Nonetheless, a clumpy torus could alter this relationship by introducing variability in the line-of-sight, as it allows encountering different \NH{} for the same orientation at different azimuthal angles. This may contribute to the observed scatter and could be explored in future work.

\subsection{Intrinsic fraction of Compton-thick AGN} \label{subsec:ctcontribution}

\begin{figure*}[t]
\subfloat{\includegraphics[width = 0.5\hsize]{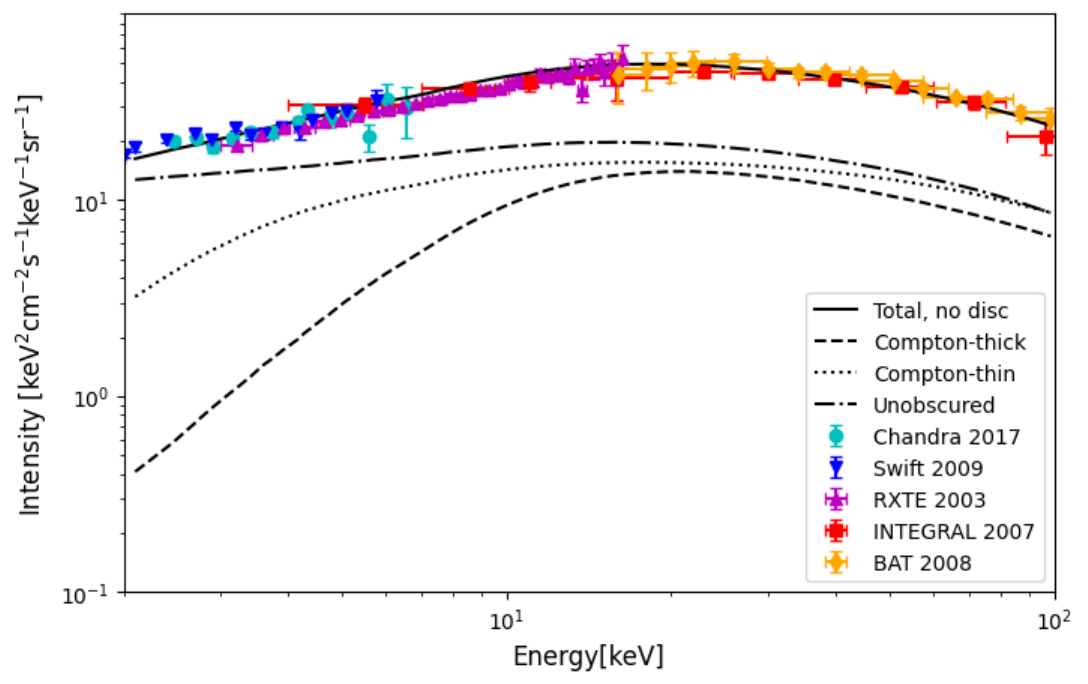}} 
\subfloat{\includegraphics[width = 0.5\hsize]{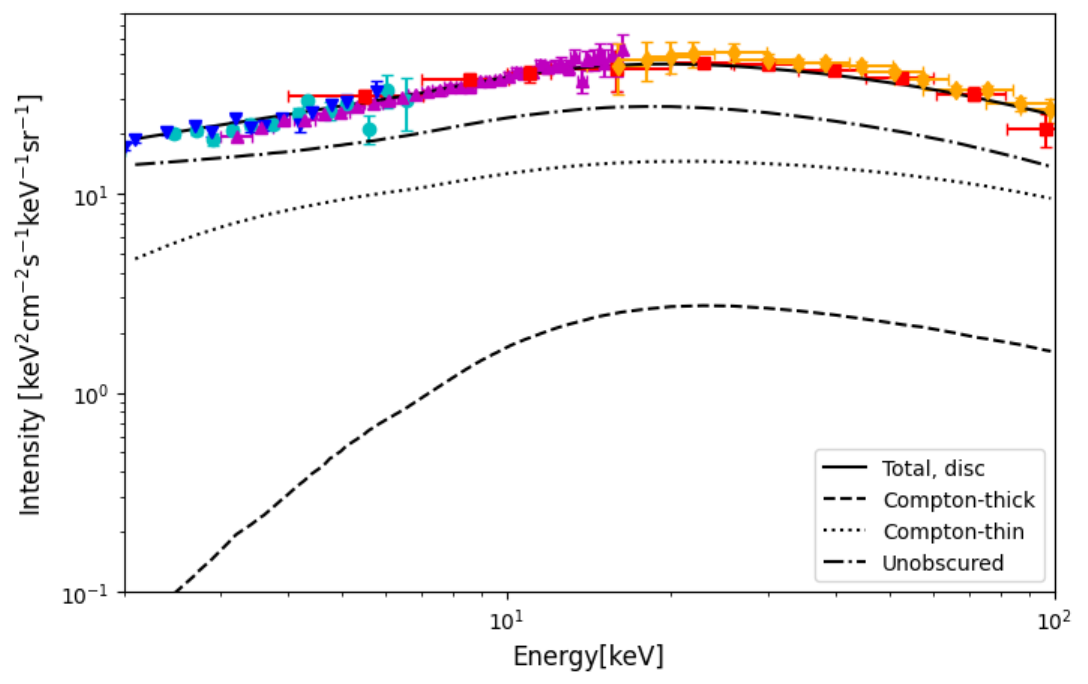}}
\caption{\textit{Left:} Contribution of unobscured (log(\NH{}/\,cm$^{-2}$ < 22; dashed-dotted line), Compton-thin (22 $\leq$ log(\NH{}/\,cm$^{-2}) <$ 24; dotted line) and CT sources (log(\NH{}/\,cm$^{-2}) \geq$ 24; dashed line) to the CXB  (solid line) reproduced by the median parameters of the LD torus model. \textit{Right:} As before but for the LD+AD model. \label{fig:cxbcontribution}}
\end{figure*}

We derive the intrinsic CT fraction (log(\NH{}/\,cm$^{-2}) \geq$ 24) with the LD and LD+AD models and find that it drops from 50$\pm$2\% (68\% confidence based on the posteriors) without the AD to 21$\pm$7\% with the AD. As shown in Fig. \ref{fig:ref-abs}, when adding the AD, the unobscured and Compton-thin AGN produce more reflection without additional absorption, reducing the amount of CT AGN needed to reproduce the CXB peak. As the LD+AD model is more realistic and reproduces best all constraints, the CT fraction of 21$\pm$7\% is likely closer to the true value. This aligns very well with the findings of local X-ray surveys \citep[e.g.,][]{2011ApJ...728...58B, 2015ApJ...815L..13R, 2025ApJ...978..118B} and with the predictions of previous AGN population syntheses at low redshift \citep[e.g.,][]{2014ApJ...786..104U, 2015MNRAS.451.1892A}. However, at $z$ = 1, \cite{2014ApJ...786..104U} and \cite{2019ApJ...871..240A} predict an intrinsic CT fraction of 37$\pm$2\% and 56$\pm$7\%, respectively. The discrepancy with our value likely arises at least in part because our model does not take into account redshift evolution, causing the CT fraction of the parent population to remain constant. This limitation may also explain why our model struggles to accurately match the absorbed AGN fraction at higher redshifts (but see Sect. \ref{subsec:galacticplane} for a different explanation). To incorporate evolution in our model, an additional parameter would be needed to evolve the size or density of the dusty torus, which is beyond the scope of this work.

\begin{figure}[t]
\subfloat{\includegraphics[width = \hsize]{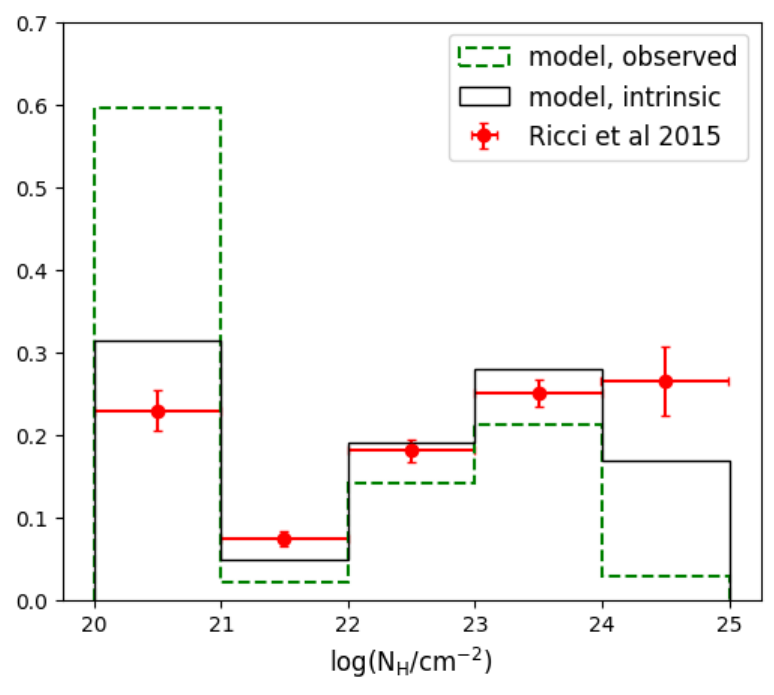}} 
\caption{Intrinsic \NH{} distribution (black line) of our local synthetic population produced by the LD+AD model with the combined constraints. Our observed distribution (green line) from Fig. \ref{fig:recdisccombinedresults} and the intrinsic distribution from \cite{2015ApJ...815L..13R} (red points) are shown for comparison. \label{fig:intrinsicnh}}
\end{figure}

In Fig. \ref{fig:cxbcontribution}, we further compare the contribution of unobscured [log(\NH{}/\,cm$^{-2}$) < 22], Compton-thin [22 $\leq$ log(\NH{}/\,cm$^{-2}) <$ 24] and CT sources [log(\NH{}/\,cm$^{-2}) \geq$ 24] to the CXB with the LD and LD+AD models. The total contribution of CT objects to the CXB in the 14-195 keV range is found to be $\sim28\%$ for the LD model, while the CT contribution drops to about 7\% with the LD+AD model. The latter value is in close agreement with that found by \cite{2009ApJ...696..110T}.

Finally, in Fig. \ref{fig:intrinsicnh}, we present the intrinsic line-of-sight \NH{} distribution of our local population of AGN using the LD+AD model. The distribution reveals the same bimodality seen in the observed one, with peaks in the log(\NH{}/\,cm$^{-2}$) = 20--21 and 23--24 bins. Our model recovers well the results from \cite{2015ApJ...815L..13R}, although they assume a fixed covering factor when correcting the observed \NH{} distribution into the intrinsic one. For reference, we also include the observed distribution of our model from Fig. \ref{fig:recdisccombinedresults}, with the first two bins separated. The selection bias is most significant for completely unobscured and CT AGN, as expected.

\subsection{Origin of mildly obscured AGN} \label{subsec:galacticplane}
As discussed in Sect. \ref{subsec:obscurationobs}, our models do not contain the ingredients to generate a population with a significant fraction of mildly obscured AGN [log(\NH{}/\,cm$^{-2}$) < 22]. One possible origin of obscuration in these AGN is the gas and dust in the plane of the host galaxy around our dusty torus, that would cause completely unobscured objects to be subject to additional absorption of log(\NH{}/\,cm$^{-2}$) < 22. \cite{2022A&A...666A..17G} construct a clumpy interstellar medium (ISM) whose characteristic cloud surface density evolves with redshift, and use it in combination with the obscuring torus to reproduce the fraction of absorbed AGN as a function of redshift from deep X-ray studies \citep[e.g.,][]{2015ApJ...802...89B, 2020A&A...639A..51I}. They find that the median \NH{} of the ISM evolves as
$\propto (1 + z)^{3.3}$, assuming a median \NH{} = $10^{21}$ cm$^{-2}$ for local galaxies \citep{2013MNRAS.431..394W}.

Hence, if we incorporate this component into our model, the ISM \NH{} could increase by up to $\sim2.4$ times within our local ($z < 0.3$) synthetic population, which is used to reproduce the observed fraction of \NH{} in bins of log(\NH{}). This increase would be sufficient to shift a significant fraction of completely unobscured objects into the log(\NH{}/\,cm$^{-2}$) = 21--22 bin. Furthermore, in the redshift range $z = 1.2 - 3.2$, the ISM \NH{} would increase from $\sim13$ to more than 100 times. This could result in the increase of the absorbed fraction at higher redshift, which our current model underpredicts. Regarding our model parameters, incorporating the ISM would likely cause the posterior of the mean, $\mu$, of the \NHeq{} distribution to shift to lower values. Our results on the CXB would likely not be significantly affected, since the biggest difference would be high redshift sources that are obscured (see right of Fig. \ref{fig:cxbcontribution}).

\section{Conclusions} \label{sec:conclusions}
Until now, population synthesis models of AGN have not treated absorption and reflection self-consistently, but rather assumed values for the reflection at different obscurations and for the intrinsic fraction of CT AGN. We have created a fully self-consistent population synthesis model of AGN with numerical simulations of their X-ray emission using the ray-tracing code \textsc{RefleX}. Our constraints include the CXB as well as absorption properties of AGN detected in different X-ray surveys, namely the fraction of \NH{} in bins of log(\NH{}) and the fraction of absorbed AGN as a function of observed $L$\textsubscript{X} and as a function of redshift. We sample an intrinsic XLF and parameterise the geometry and density of the circumnuclear material of the AGN population. We construct geometrical models based on the simple unification of AGN and the receding torus, and lastly include an AD. We determine the parameter posterior distributions for each model using an SBI method.

We find that the CXB can be well reproduced even by the simple torus model (Fig. \ref{fig:cxbresults}). However, matching the absorption properties requires the LD torus model (Fig. \ref{fig:absresults}). The LD+AD model best satisfies all observational constraints simultaneously (Fig. \ref{fig:recdisccombinedresults}). In this case, the posterior distribution of the slope in Eq. (\ref{eq:RoutLx}) has a median of $\alpha$ = 0.24$\mathrm{_{-0.01}^{+0.01}}$, agreeing with X-ray observations.

We derive an intrinsic CT fraction of 21$\pm$7\% (68\% confidence based on the posteriors), in agreement with observations of the local Universe, and find that the inner edge of the dusty torus spans a range of 0.2--2.8\,pc in the L$\mathrm{_X^{int}}$ range 10$^{41} - 10^{46}$ erg s$^{-1}$, consistent with dust sublimation radius sizes from infrared interferometry studies. Our synthetic population can reproduce the luminosity dependence of the torus covering factor, XMM-\textit{Newton}, \textit{Chandra}, and NuSTAR number counts, as well as the observed correlation between reflection and obscuration (Figs. \ref{fig:coveringfactor}, \ref{fig:counts}, \ref{fig:ref-abs}). Additionally, it can recover the XLFs of absorbed and unabsorbed AGN, and the intrinsic \NH{} distribution of local AGN. However, our AGN modelling lacks the obscuration originating at the host galaxy as well as evolution, resulting in a lower fraction of absorbed AGN at high redshifts and a constant CT fraction.

This is the first work in which the CXB is reproduced with a fully self-consistent synthetic AGN population model, without assuming the reflection features of the spectra, the intrinsic \NH{} distribution or the CT fraction, and without modifying the XLF. 
Modelling and linking the emission, absorption, and reflection through the matter distribution allows us to break the degeneracy of fitting the CXB by simultaneously matching observational constraints concerning absorption, which has not been done before. Overall, our results emphasise that a better understanding of AGN requires models that account for both the geometric structure and the physical mechanisms of the absorbers in order to explain their observed properties. In the future, this methodology can be extended to include more complex models that can account for effects such as clumpiness of the obscuring material, evolution, or additional relationships between AGN parameters, introducing in particular the intrinsic properties of SMBHs, such as their mass and Eddington ratio.

\begin{acknowledgements}
      CR acknowledges support from Fondecyt Regular grant 1230345, ANID BASAL project FB210003 and the China-Chile joint research fund.
\end{acknowledgements}

          
\bibliographystyle{aa}
\bibliography{Bib}

\begin{appendix}
\section{SBI posteriors} \label{app:sbiposteriors}

\begin{figure}[ht]
\centering
\subfloat{\includegraphics[width =1.\hsize]{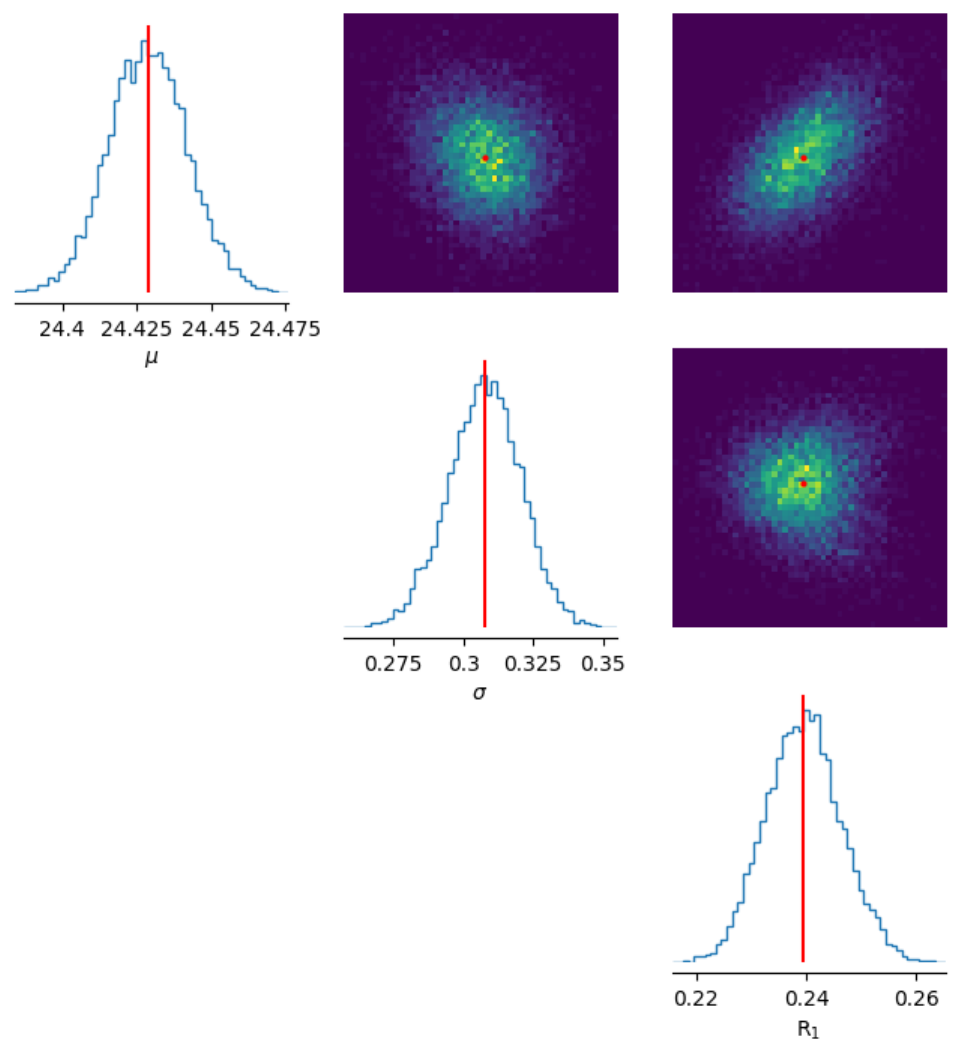}}
\caption{Parameter posteriors using the simple torus model when applied on the CXB data sets. The median of each distribution reported in Table \ref{table:simpletorusfits} is denoted in red.
\label{fig:cxbpost}}
\end{figure}

\begin{figure*}[ht]
\centering
\subfloat{\includegraphics[width =0.5\hsize]{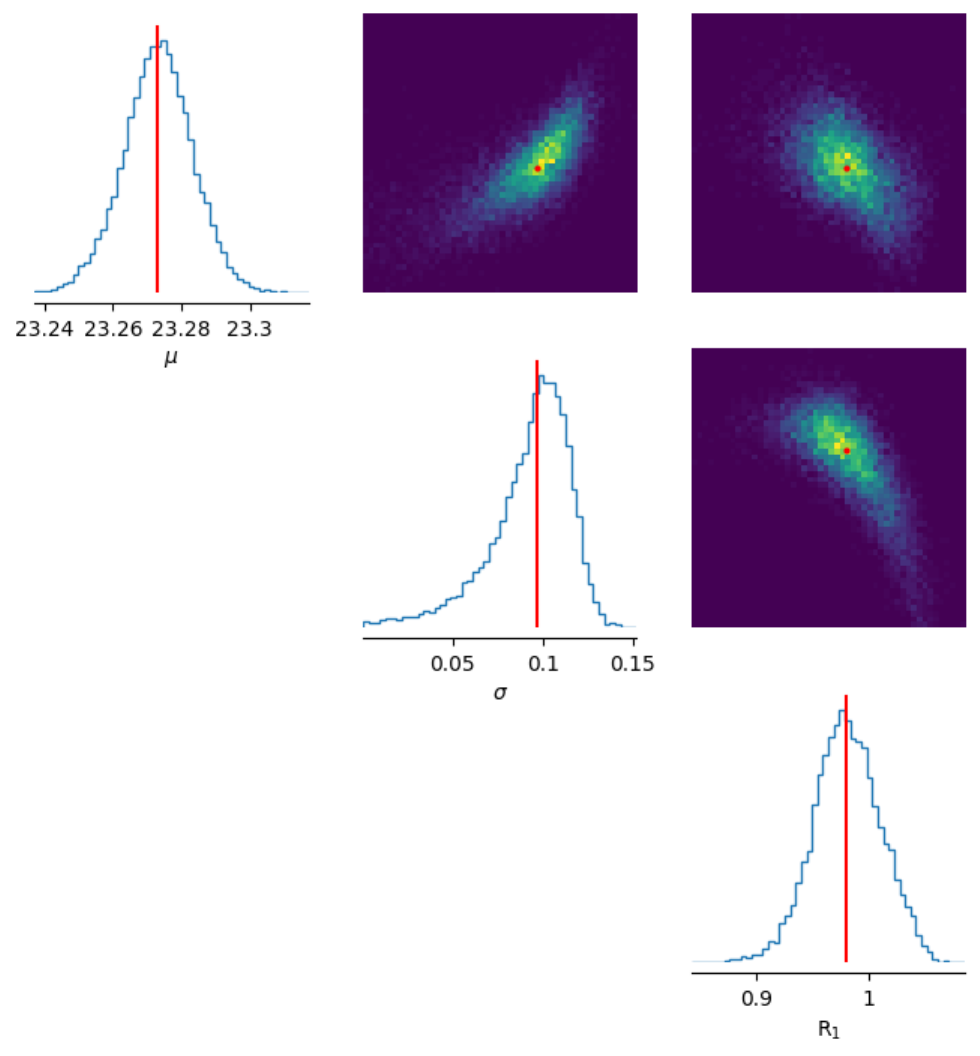}}
\subfloat{\includegraphics[width =0.5\hsize]{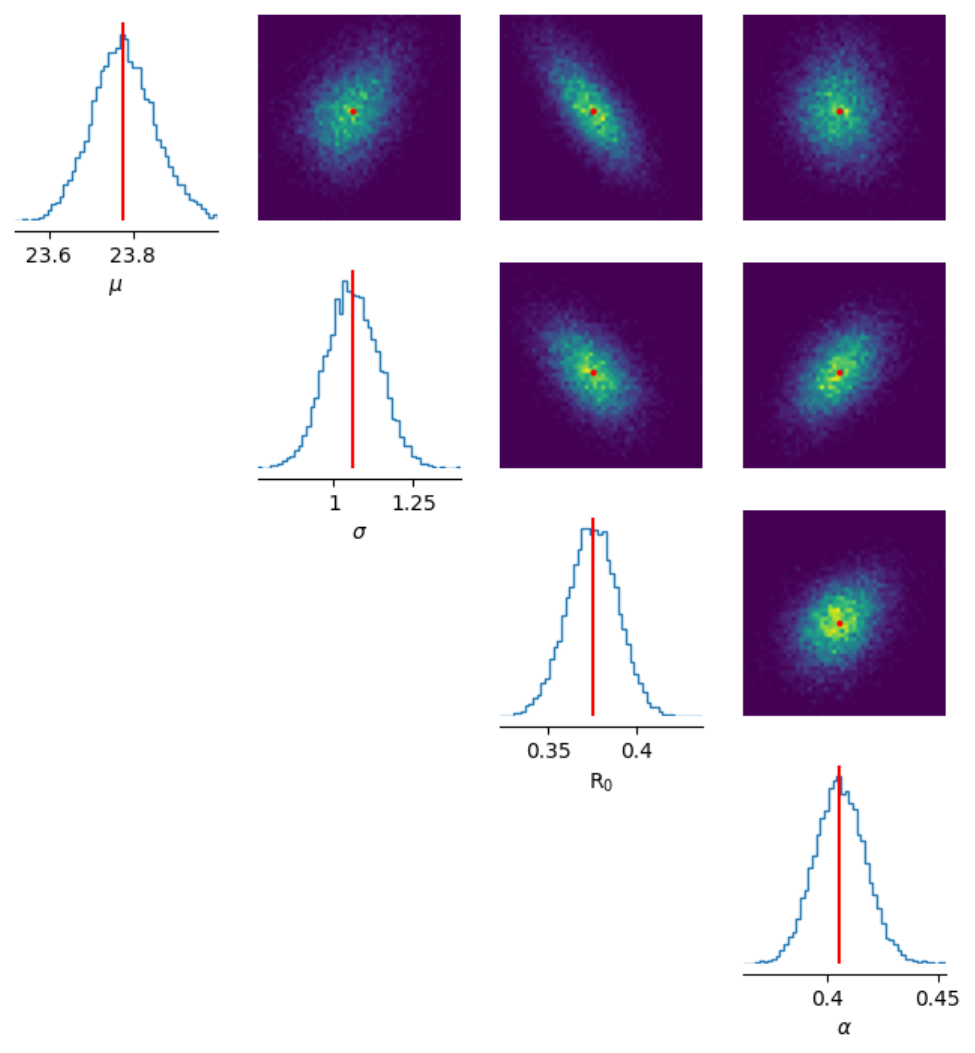}}
\caption{Parameter posteriors using \textit{left:} the simple torus model and \textit{right:} the LD torus model, when applied on the absorption properties. The median of each distribution reported in Table \ref{table:simpletorusfits} is denoted in red.
\label{fig:abspost}}
\end{figure*}

\begin{figure*}[ht]
\centering
\subfloat{\includegraphics[width =0.5\hsize]{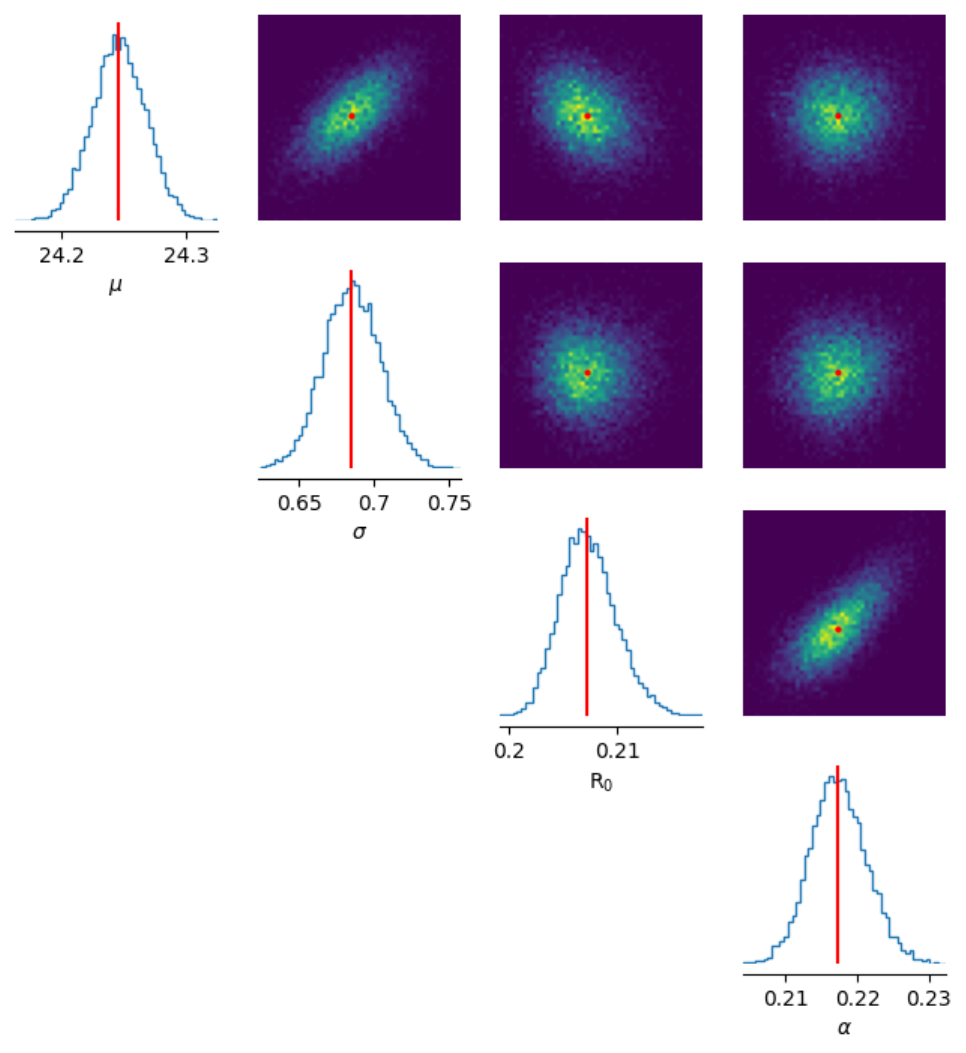}}
\subfloat{\includegraphics[width =0.5\hsize]{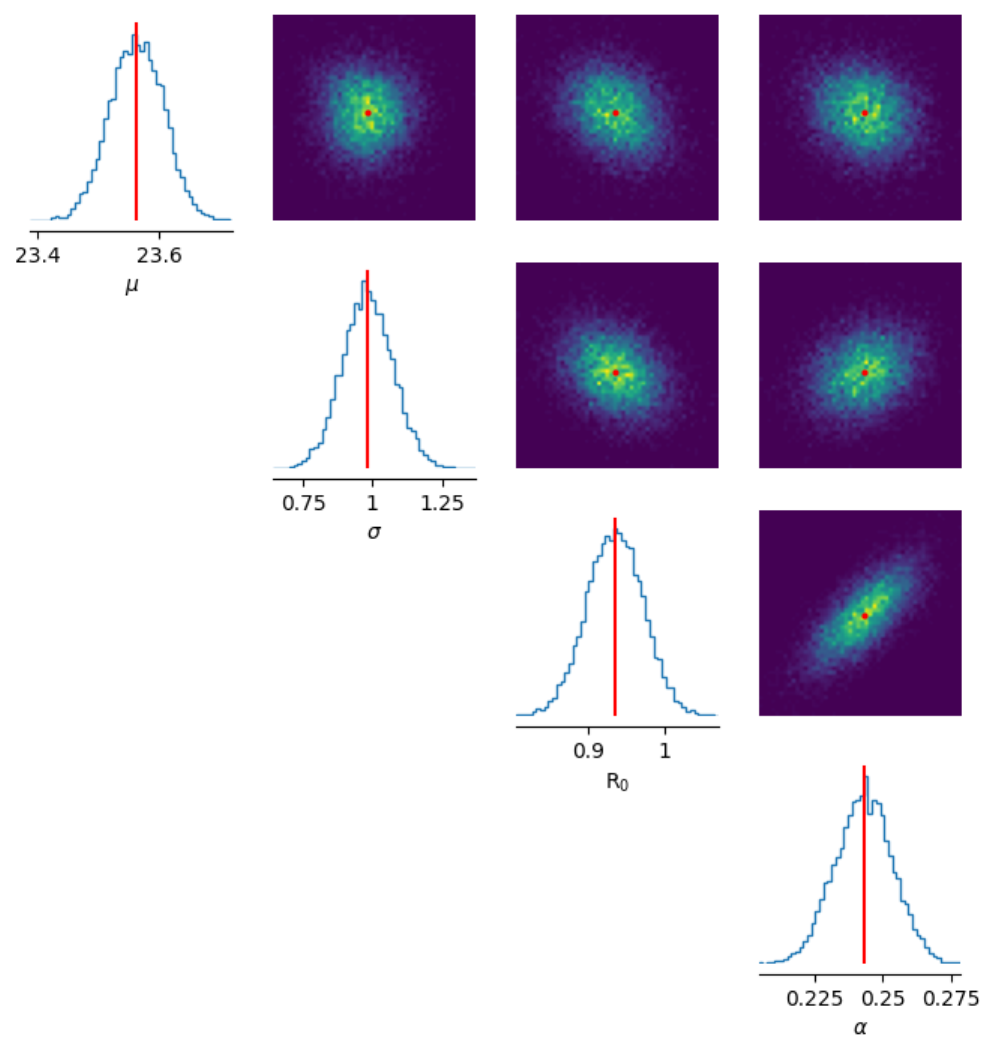}}
\caption{Parameter posteriors using \textit{left:} the LD torus model and \textit{right:} the LD+AD model, when applied on all constraints. The median of each distribution reported in Table \ref{table:simpletorusfits} is denoted in red.
\label{fig:totalpost}}
\end{figure*}

\end{appendix}

\label{LastPage}
\end{document}